\documentclass[preprint,aps]{revtex4-1}
\usepackage{graphicx}
\usepackage{amsmath}
\usepackage{hyperref}
\usepackage{epstopdf}
\usepackage{txfonts}
\usepackage{float}
\usepackage{slashed}
\usepackage{color}
\def\be{\begin{equation}}
\def\ee{\end{equation}}
\def\bea{\begin{eqnarray}}
\def\eea{\end{eqnarray}}
\def\ba{\begin{array}}
\def\ea{\end{array}}
\def\bc{\begin{center}}
\def\ec{\end{center}}
\begin{document}
\title{Transverse densities and generalized parton distributions of $\rho$ meson in light front quark model}
\author{Narinder Kumar}
\affiliation{Department of Physics,\\ Indian
Institute of Technology Kanpur,\\ Kanpur-208016, India}
\begin{abstract}
We have investigated the transverse charge density for longitudinal and transversely polarized $\rho$ meson in light-front quark model (LFQM). Charge densities are obtained from the elastic form factors of the $\rho$ meson calculated in LFQM including the zero-mode contributions. We have computed the different helicity matrix elements of the $\rho$ meson. In addition to this, we have also presented the results for the generalized parton distributions (GPDs) and impact-parameter dependent parton distribution functions (ipdpdfs) of the $\rho$ meson.
\end{abstract}
\maketitle
\section{Introduction}
Electromagnetic form factors (FFs) are the key source to understand the internal structure of the hadrons. By taking the Fourier transform of the form factors of hadrons one can get the spatial distributions like charge and magnetization densities \cite{Carlson:2007xd,Miller:2007uy}.  These densities provide the model independent information about the charge distribution. Basically, it gives the density of hadron at transverse distance from the center of momentum.  It is to be noted that information about the charge density in terms of longitudinal coordinates is not obtained in a meaningful way. Principally, hadron states have explicit dependence on momentum while initial and final hadron states have different momentum which invalidates their interpretation as probability density. However, it is shown in Ref. \cite{Miller:2010nz} that charge density can be calculated from the hadronic form factor  which directly involve the 3D charge density of partons.  For the case of nucleons there have been considerable efforts made to understand these densities \cite{Kim:2008ghb,Kumar:2014coa,Mondal:2016xpk,Mondal:2016xsm,Chakrabarti:2014dna,Tiator:2008kd} in various phenomenological models \cite{Vega:2010ns,Vega:2012iz,Selyugin:2009ic,Granados:2013moa}. Transverse densities in fixed light-front time has been studied in Ref. \cite{Miller:2009sg,Venkat:2010by}. One of the most important work in Ref. \cite{Miller:2007uy}  showed that neutron charge density is negative at the center of core and also spatial extent of magnetization density for the proton is much larger than that of its charge density \cite{Miller:2007kt}.\\
Except nucleons, a lot of work has been done to obtain the charge densities of spin-1 system like deuteron \cite{Carlson:2008zc,Mondal:2017lph,Huang:2017gih}. For a complete overview over the theoretical and experimental studies on the deuteron form factors, see Ref. \cite{Abbott:2000ak,Garcon:2001sz,Kohl:2008zz,Holt:2012gg}. Despite spin-1 system like deuteron, $\rho$ meson is also an interesting subject to investigate. $\rho$ meson is a short-living hadronic particle having three states denoted as $\rho^+, \rho^0$ and $\rho^-$. Due to spin-1, there are three form factors charge $(G_C)$, magnetic $(G_M)$ and quadrupole $(G_Q)$ respectively for $\rho$ meson. $\rho$ meson is the lightest strong interacting particle after pions and kaons. From experimental point of view, diffractive photo production  and electroproduction of the $\rho$ meson is extensively studied at HERA H1 experiment \cite{Aktas:2006qs,Aaron:2009xp,Adloff:1999kg,Crittenden:2001tn,Aid:1996bs,H1:2015bxa,Aaron:2009wg}. In addition to this, $\rho$ meson light-front wavefunctions (LFWFs) is  also extracted using HERA data on diffractive $\rho$ photoproduction \cite{Forshaw:2010py}.  Space-like and time-like $\gamma^* \pi \rightarrow \rho$ and $\gamma^* \rho \rightarrow \pi$ transition form factors are also studied in light-cone formalism \cite{Yu:2007hp}. Recently, Sun {\it et al.}  have also discussed the generalized parton distributions and parton distributions in impact-parameter for $\rho$ meson \cite{Sun:2017gtz,Sun:2018tmk}.\\ 
Recently, 
AdS/QCD formalism has been successfully applied to various hadronic properties like, generalized parton distributions (GPDs), parton distribution functions (PDFs), form factors and transverse densities \cite{Abidin:2008ku,Brodsky:2006uqa,Brodsky:2007hb,Erlich:2005qh,SS,AC,ads1,ads101,Model,Model1,ads2,BT1,Grigoryan:2007my,BT2,deTeramond:2005su,vega,vega01, cM,CM2,CM3,HSS,Mondal,Ma_baryon,reso,Chakrabarti:2016lod,BT_reso,BT_new1,BT_new2,BT_new3,deTeramond:2013it,Sufian:2016hwn}. Regarding spin-1 particle like $\rho$ meson, a lot of work has been done in holographic QCD \cite{Forshaw:2012mb,Forshaw:2013oaa,Ahmady:2016ujw,Forshaw:2012im,Forshaw:2011yj}.\\
In the present work, we have studied the transverse charge densities for unpolarized and polarized $\rho$ meson in light-front quark model (LFQM) \cite{Jaus:1989au,Jaus:1991cy,Jaus:1996np,Melikhov:1996ge,Melikhov:1995xz,Choi:1996mq,Choi:1997qh,Choi:1997iq}. This model is quite successful in explaining the various electroweak properties of light and heavy mesons compared with experimental data. A very important calculations on the vector meson form factor (e.g. $\rho$ meson) is discussed in Ref. \cite{Bakker:2002mt}. Calculations on the structure of wave functions of mesons as bound state on relativistic quarks are discussed in Ref. \cite{Terentev:1976jk}. Form factors of the $\rho$ meson in light-front constituent quark model are also discussed in Ref. \cite{Cardarelli:1994yq}. Further, time-like form factor of $\rho$ meson are calculated in Ref. \cite{deMelo:2016lwr}.  Distribution amplitudes, decay constant and radiative decays for mesons have been also studied in Ref. \cite{Choi:2007yu} and \cite{DeWitt:2003rs} respectively. In addition to this, skewed parton distributions for pion have been also discussed in LFQM \cite{Choi:2001fc}. \\
Transverse charge densities are obtained by taking the two-dimensional Fourier transform of the physical form factors of $\rho$ meson. Transverse densities are also related with the generalized parton distributions (GPDs) with zero skewness \cite{Burkardt:2000za,Burkardt:2002hr}. Although GPDs reveals simultaneously information on both longitudinal and the transverse distribution of partons in a fast moving hadron. This physical picture becomes more intuitive when one take a Fourier transform from transverse momentum transfer to impact-parameter space. GPDs can be accessed by deep virtual Compton scattering process (DVCS) and deep virtual meson production(DVMP) \cite{Boffi:2007yc}. In this work, we have also calculated the GPDs and impact-parameter dependent parton distribution function (ipdpdf) for $\rho$ meson \cite{Diehl:2002he,Burkardt:2002ks}.\\
To complete this work, we have used the results of physical electromagnetic form factors of $\rho$ meson computed by Choi {\it et al.} in Ref.  \cite{Choi:2004ww} by including the zero-mode contribution \cite{Chang:1973qi,Burkardt:1989wy,deMelo:1998an,Brodsky:1998hn}. Authors take care of zero-mode issue in calculating the form factors of $\rho$ meson using different helicity components. The zero mode has close relation with the off-diagonal elements in the Fock state expansion of the current matrix. The existence of zero-mode contribution to the form factor is characterized by the contribution from off-diagonal elements in $q^+ \rightarrow 0$ limit. One can also represent the hadronic FFs without the presence of zero-mode contribution. However for spin-1 system, FFs include the zero-mode contribution which arises in the matrix element of plus current.\\
The manuscript is arranged as follows. After introduction, we discuss the physical form factors and light-front quark model in section \ref{sec:2}. In section \ref{sec:3}, we discuss the transverse charge density for $\rho$ meson. After this, we discuss the GPDs for $\rho$ meson in section \ref{sec:4} followed by ipdpdfs in section \ref{sec:5}. Conclusions are drawn in last section.

\section{Physical form factors and light-front quark model}
\label{sec:2}
In the present work, we have used the LFQM which successfully explain various electromagnetic properties of the hadrons. For the spin-1 particle, Lorentz invariant electromagnetic form factors $(F_i (i=1,2,3))$ are defined by the matrix elements of the $J^\mu$ current sandwiched between initial $| P, \Lambda \rangle$ and final $\langle P', \Lambda'|$ states as follows \cite{Choi:2004ww}:
\begin{eqnarray}
\langle P', \Lambda' | J^\mu | P, \Lambda \rangle &= & - \epsilon_{\Lambda'}^{*} \cdot \epsilon_\Lambda (P+P')^\mu F_1(Q^2) + ( \epsilon_{\Lambda'}^{\mu} q \cdot \epsilon_{\Lambda'}^{*} 
- \epsilon_{\Lambda'}^{* \mu} q \cdot \epsilon_\Lambda ) F_2 (Q^2) + \nonumber\\
&& \frac{(\epsilon_{\Lambda'}^{*} \cdot q)(\epsilon_\Lambda \cdot q)}{2 M_v^2} (P+P')^\mu F_3(Q^2),
\label{electromagnetic}
\end{eqnarray}
where $q=P'-P$ and $\epsilon_\Lambda (\epsilon_\Lambda') $ is the polarization vector of the initial (final) meson with the physical mass $M_v$.  We employ the Breit frame \cite{Cardarelli:1994yq,Choi:1996mq,Bakker:2002mt} $(q^+=0, q_y=0, q_x=Q, {\bf P}_\perp = - {\bf P'}_\perp)$ and the momenta are defined by 
\begin{eqnarray}
q^\mu &=& (0,0, Q, 0), \nonumber\\
P^\mu &=& (M_v \sqrt{1+\kappa}, M_v \sqrt{1+\kappa}, -Q/2,0), \nonumber\\
P'^\mu &=& (M_v \sqrt{1+\kappa}, M_v \sqrt{1+\kappa}, Q/2,0),
\end{eqnarray}
and $\kappa = Q^2/ 4 M_v^2$ and the notation used is $p^\mu = (p^+, p^-, p^1, p^2)$. One can obtain the covariant form factors of spin-1 particle by using the plus component of the current $I_{\Lambda' \Lambda}^{+}(0) = \langle P', \Lambda' | J^+ | P, \Lambda \rangle$. It is also clear from Eq. (\ref{electromagnetic}), that their are three form factors and total nine elements of $I^{+}_{\Lambda' \Lambda}(0)$ can be assigned to the current operator. Due to light-front parity and light-front time reversal invariance, we can reduce it to only four elements, $I_{++}^{+}, I_{+-}^{+}, I_{+0}^{+} $ and $I_{00}^{+}$. Practically, the physical charge $(G_c)$, magnetic $(G_M)$ and quadrupole $(G_Q)$ form factors are often used to describe electromagnetic properties of hadron rather than $F_i(Q^2) (i=1,2,3)$ form factors. The relation between these two form factors are given by:
\begin{figure*}
\begin{minipage}[c]{0.98\textwidth}
    \small{(a)}\includegraphics[width=7cm]{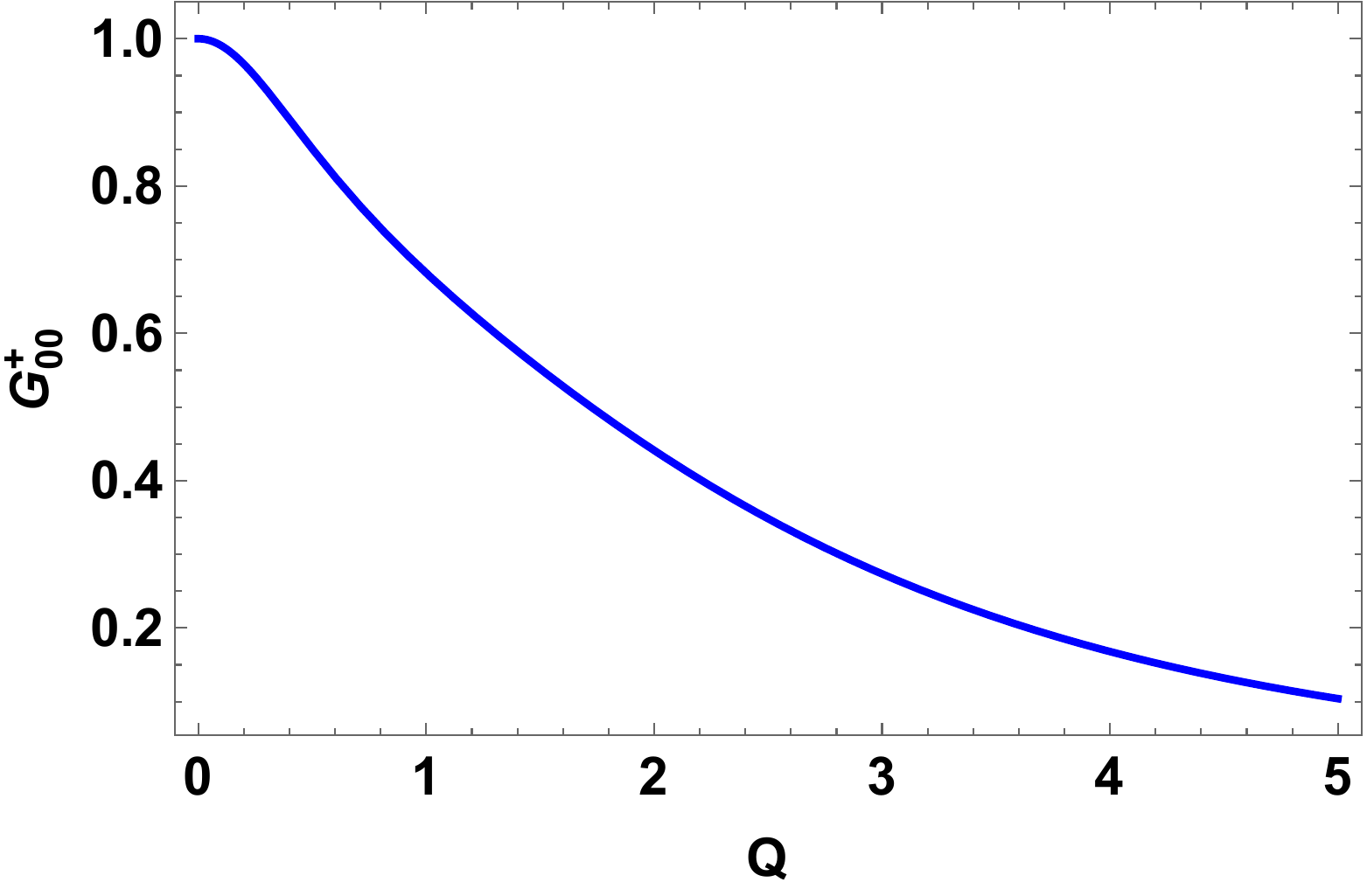}
    \hspace{0.1cm}
    \small{(b)}\includegraphics[width=7cm]{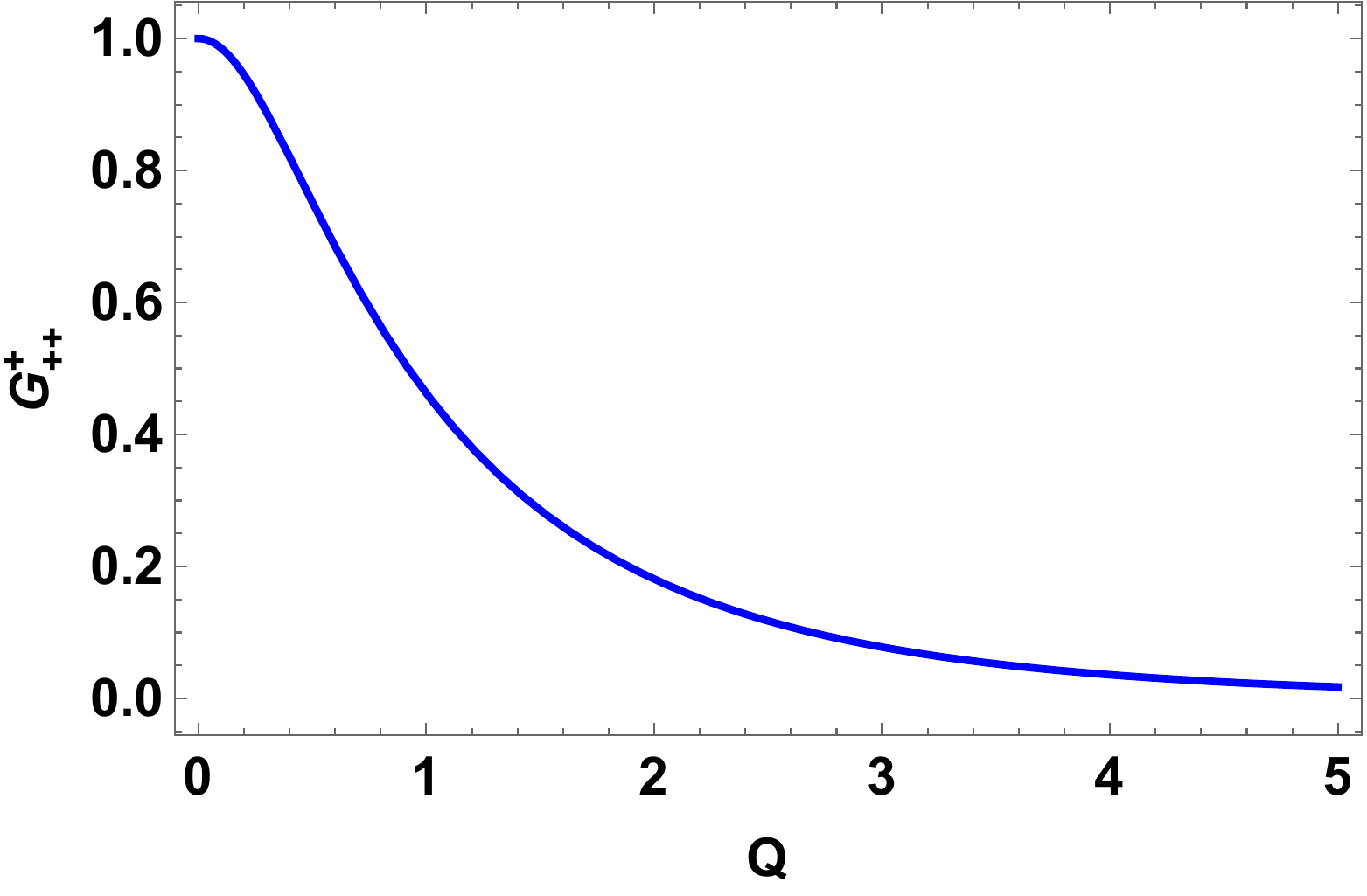}
    \hspace{0.1cm}
    \end{minipage}
\begin{minipage}[c]{0.98\textwidth}
  \small{(c)}   \includegraphics[width=7cm]{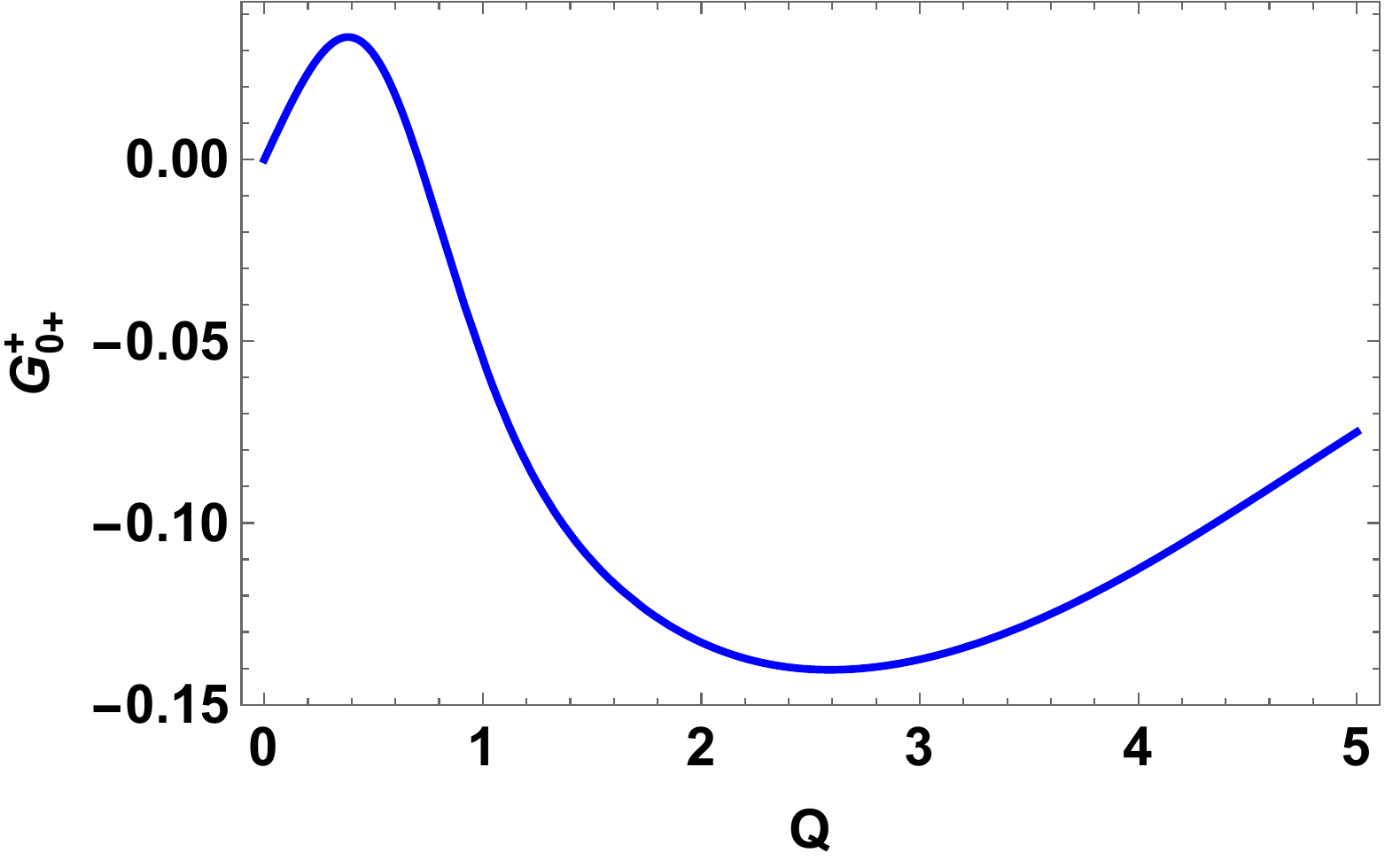}
  \hspace{0.1cm}
   \small{(d)}  \includegraphics[width=7cm]{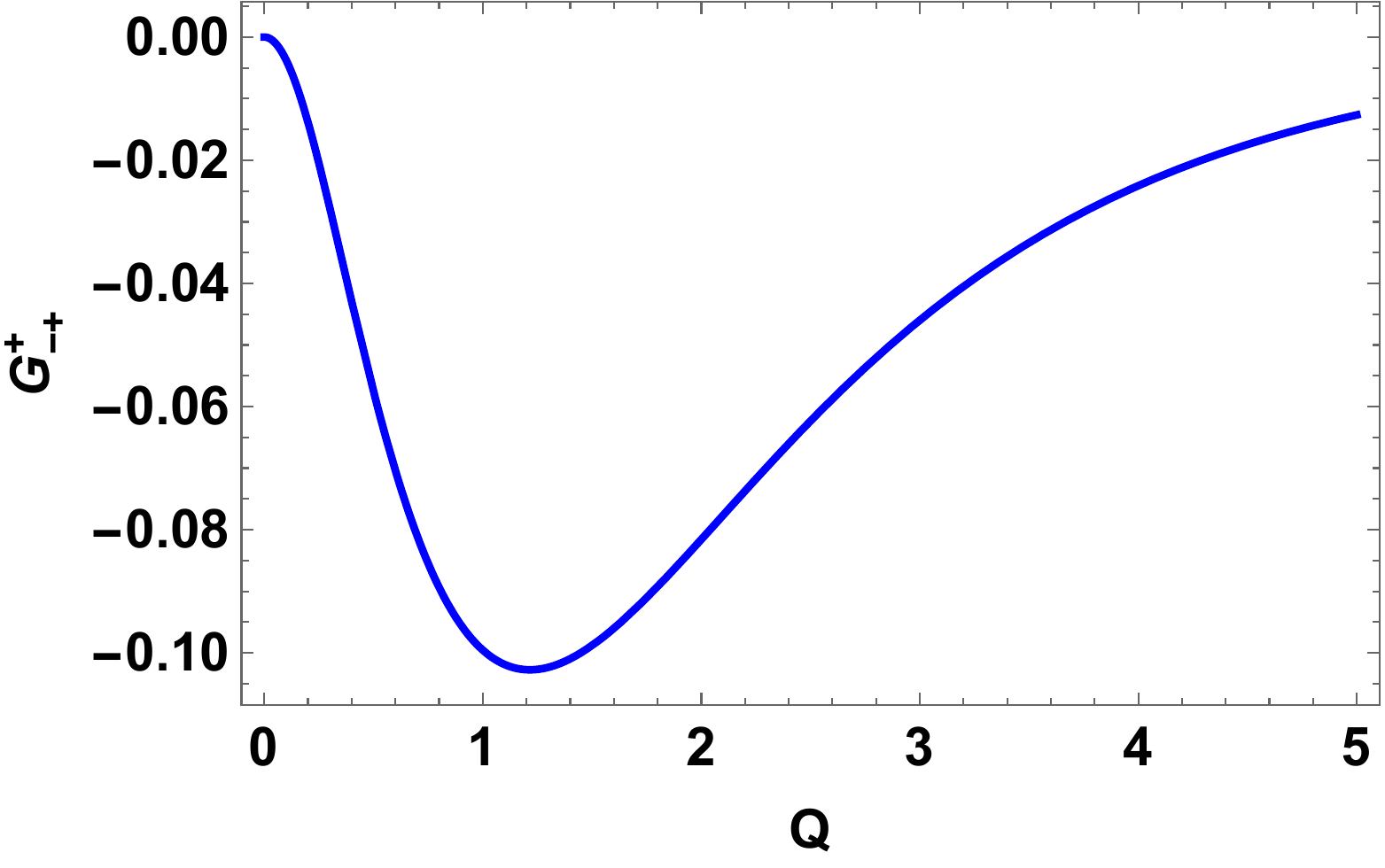}
     \hspace{0.1cm}
\end{minipage}
\caption{(Color online) Plots of helicity-conserving matrix elements (a) $G_{00}^{+}(Q^2)$, (b) $G_{++}^{+}(Q^2)$ and (c) helicity-flip matrix element with one unit of helicity-flip $G_{0+}^{+}(Q^2)$ and (d) with two unit of helicity-flip $G_{-+}^{+}(Q^2)$.}
\label{helicity-ffs}
\end{figure*}
\begin{eqnarray}
G_C &=& F_1 + \frac{2}{3} \kappa \ G_Q, \nonumber\\
G_M &=& - F_2, \nonumber\\
G_Q&=& F_1 +F_2 +(1+\kappa) F_3.
\end{eqnarray}
In addition to this, at zero momentum transfer these form factors are equal to static charge $e$, magnetic moment $\mu$ and quadrupole moment $Q$ respectively:
\begin{eqnarray}
e \ G_c(0) &=& e,\nonumber\\
e \ G_m(0) &=& 2 \ M_v \  \mu, \nonumber\\
-e \ G_Q(0) &=& M_v^2 \ Q.
\end{eqnarray}
\begin{figure*}
\begin{minipage}[c]{0.98\textwidth}
    \small{(a)}\includegraphics[width=7cm]{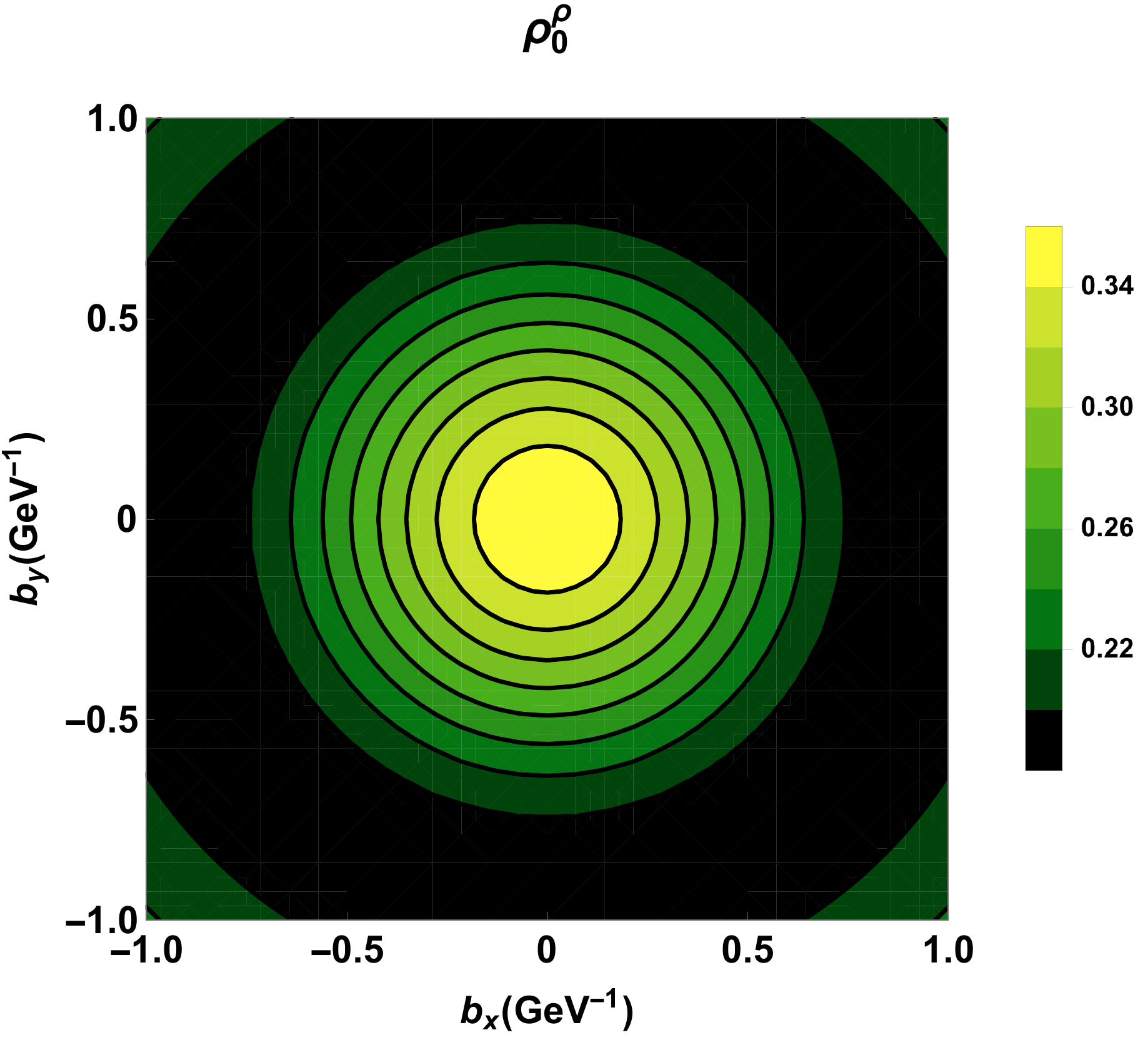}
    \hspace{0.1cm}
    \small{(b)}\includegraphics[width=7cm]{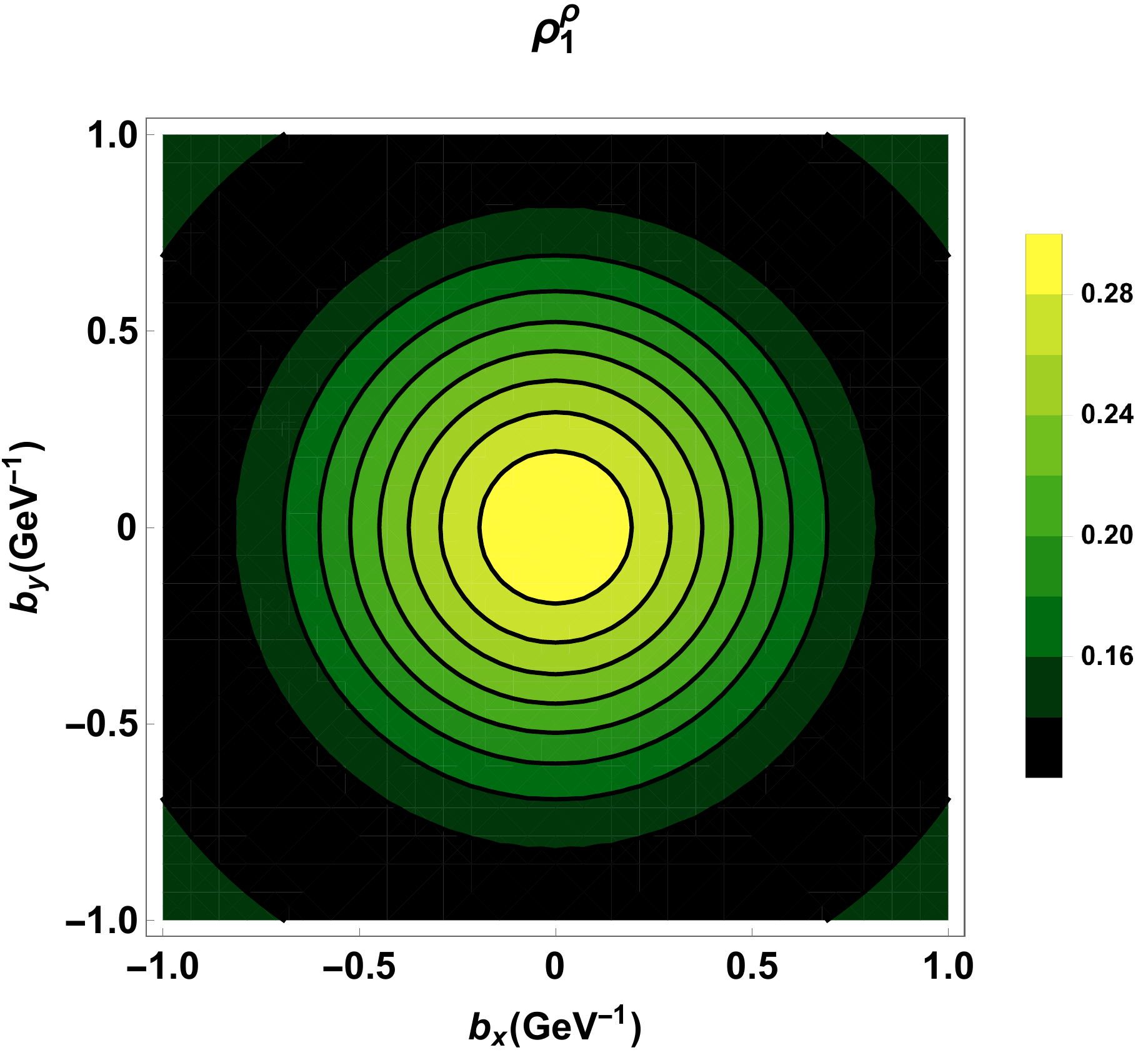}
    \hspace{0.1cm}
     \end{minipage}
     \begin{minipage}[c]{0.98\textwidth}
         \small{(c)}\includegraphics[width=7cm]{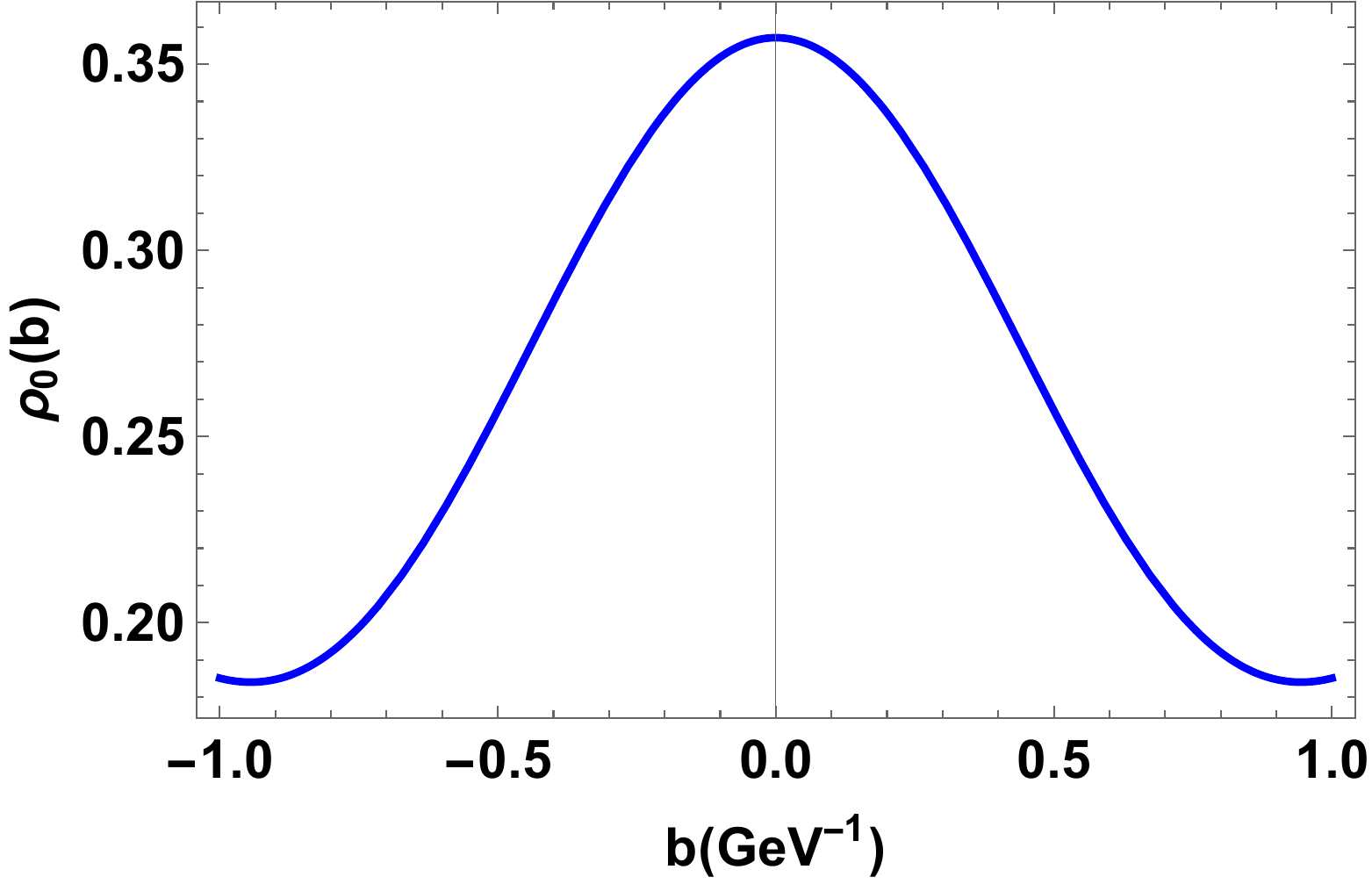}
             \hspace{0.1cm}
    \small{(d)}\includegraphics[width=7cm]{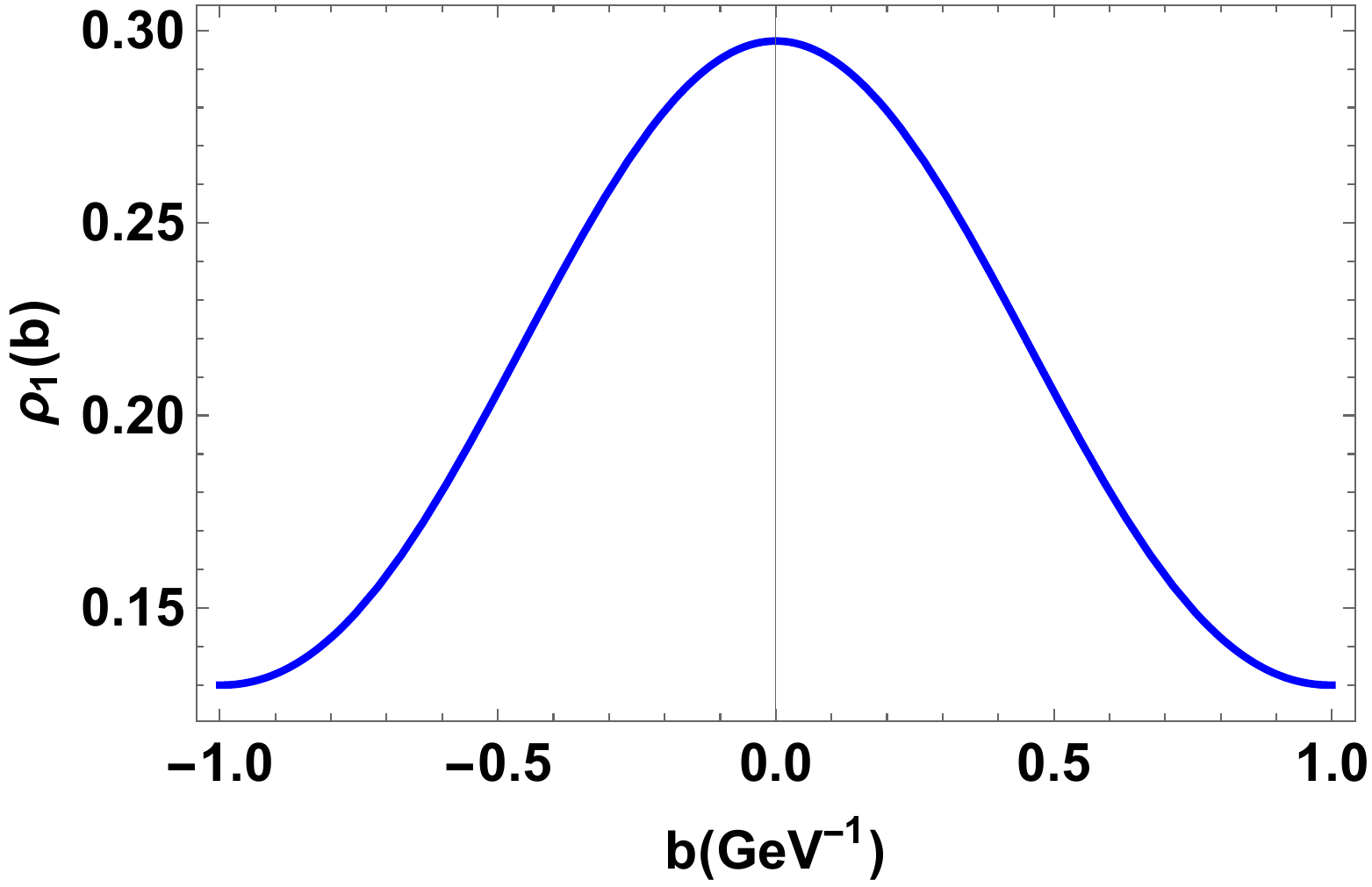}
        \hspace{0.1cm}
    \end{minipage}
\caption{(Color online) Transverse charge densities of the $\rho$ meson evaluated in light-front quark model (a) and (c) $\rho_{0}^{\rho}(b)$ and $\rho_{1}^{\rho}(b)$ for unpolarized $\rho$ meson in (b) and (d).}
\label{unpolarized-rho-meson}
\end{figure*}
For the calculation of such type of form factors Grach and Kondratyuk (GK)\cite{Grach:1983hd} and Brodsky and Hiller (BH) \cite{Brodsky:1992px} prescription has been used. However, we define the physical form factors in BH prescription which includes the zero-mode contributions and is given by
\begin{eqnarray}
G_{C}^{BH}&= &\frac{1}{2 P^+ (1+2 \kappa)} \bigg[\frac{3-2\kappa}{3} I_{00}^{+}+\frac{16}{3} \kappa \frac{I_{+0}^{+}}{\sqrt{2 \kappa}}+ \frac{2}{3} (2 \kappa -1) I_{+-}^{+}\bigg], \nonumber\\
G_{M}^{BH}&=& \frac{2}{2 P^+ (1+2 \kappa)} \bigg[I^{+}_{00} +\frac{(2 \kappa -1 )}{\sqrt{2 \kappa}} I_{+0}^{+} - I_{+-}^{+}\bigg], \nonumber\\
G_{Q}^{BH} &=& \frac{-1}{2 P^+ (1+2 \kappa)} \bigg[I^{+}_{00} - 2 \frac{I_{+0}^{+}}{\sqrt{2 \kappa}} + \frac{1+\kappa}{\kappa} I_{+-}^{+}\bigg].
\label{physical-ffs}
\end{eqnarray}
They also satisfy the angular condition which is given by
\begin{equation}
\Delta(Q^2) = (1+2 \kappa) I_{++}^{+} + I_{+-}^{+} - \sqrt{8 \kappa} \ I_{+ 0}^{+} - I_{00}^{+}.
\end{equation}
This condition must be satisfied by the matrix elements so that physical form factors must be independent from the GK or BH prescription. We can also define the structure functions $A(Q^2), B(Q^2)$ and tensor polarization $T_{20}$ by relating physical form factors as:
\begin{eqnarray}
A(Q^2)&=&G_{C}^{2}+\frac{2}{3} \kappa \ G_{M}^{2} + \frac{8}{9} \kappa^2 G_{Q}^{2},\nonumber\\
B(Q^2)&=& \frac{4}{3} \kappa \ (1+\kappa) G_{M}^{2}, \nonumber\\
T_{20} (Q^2, \theta)&=&-\kappa\frac{\sqrt{2}}{3} \frac{\frac{4}{3} G_{Q}^{2} + 4 G_Q G_C + [1/2+(1+\kappa) \tan^2(\theta/2)] G_{M}^{2}}{A+B \tan^2(\theta/2)}.
\end{eqnarray}
In LFQM, the physical form factors are obtained from the matrix element $I_{\Lambda' \Lambda}^{+}$ which is defined as
\begin{eqnarray}
I_{\Lambda' \Lambda}^{+} &=& \int \frac{dx}{2(1-x)} \int d^2{\bf k}_\perp \sqrt{\frac{\partial k'_z}{\partial x}\frac{\partial k_z}{\partial x}} \phi^*(x, {\bf k}_{\perp f})\phi(x, {\bf k}_{\perp i}) \frac{(S_{\Lambda' \Lambda}^{+})_{on}}{M_{oi} M_{of}},
\end{eqnarray}
where $(S_{\Lambda' \Lambda}^{+})_{on}$ is defined in Ref. \cite{Choi:2004ww} and radial wave function is defined as
\begin{equation}
\phi(x, {\bf k}^2)= \sqrt{\frac{1}{\pi^3/2 \beta^3}} exp(- {\bf k}^2/2 \beta^2),
\end{equation}
where ${\bf k}^2={\bf k_\perp}^2 + k_z^2$, $k_z= (x-1/2) M_o $, \cite{terent,bakker} and 
\begin{equation}
M_{oi}^2=M_{of}^2=M_o^2= \frac{{\bf k_\perp}^2 +m^2}{x(1-x)}.
\end{equation}
However, it is shown in Ref. \cite{Choi:2004ww} that zero-mode contribution arise from the $S^+_{00}$ component  which is usually avoid by considering the GK prescription over BH prescription but it is to be noted that (0,0) component is the longitudinal component  and it is the most dominant contributor in the high momentum transfer region or for the analysis in the high momentum perturbative QCD and therefore, it may be better to use the BH prescription, involving the (0,0), (+,0) and (+,-) amplitudes. \\
 It is also observed that presence of zero mode contribution in (0,0) amplitude is quite significant in light-front quark model phenomenologically because the absence of zero mode in (+,0) amplitude can give a tremendous benefit on reliable predictions on the $\rho$ meson when compared with the calculations done in the covariant formulation \cite{Jaus:1999zv, Jaus:2002sv}. It is also interesting that transverse densities are obtained from the matrix elements which feel the zero mode but they do not as model do not contain the zero mode. It is probably due to the fact that Eq. 18 described the valence contribution of $I^+_{\Lambda' \Lambda}$ and only on-shell trace terms. It is clear from Ref. \cite{Choi:2004ww} that zero mode contribution in LFQM arises from the $(S^+_{0,0})_{off}$ term. Nevertheless, in literature GK prescription is preferred over BH prescription as it is free from zero mode by definition. 
\begin{figure*}
\begin{minipage}[c]{0.98\textwidth}
    \small{(a)}\includegraphics[width=7cm]{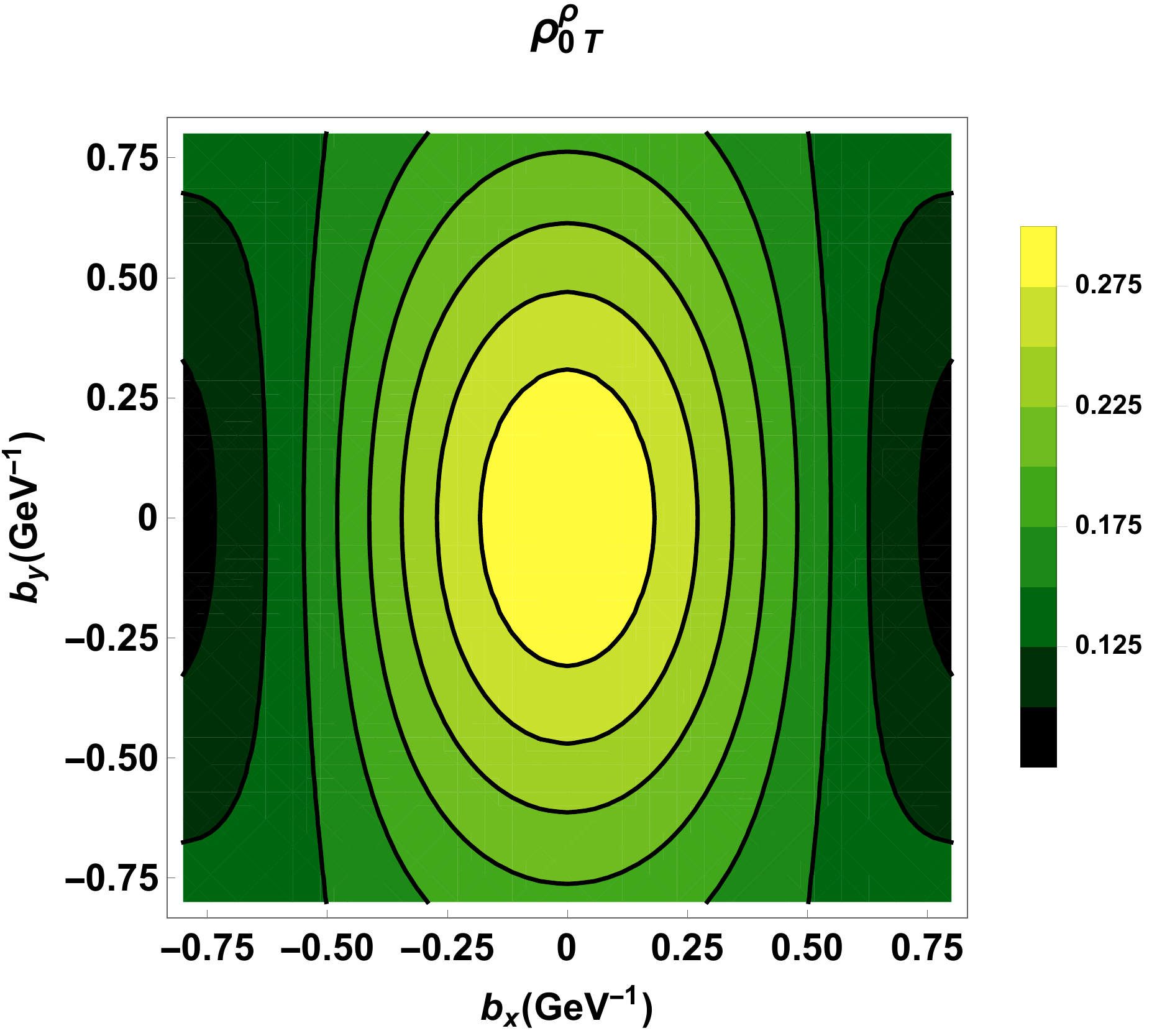}
    \hspace{0.1cm}
    \small{(b)}\includegraphics[width=7cm]{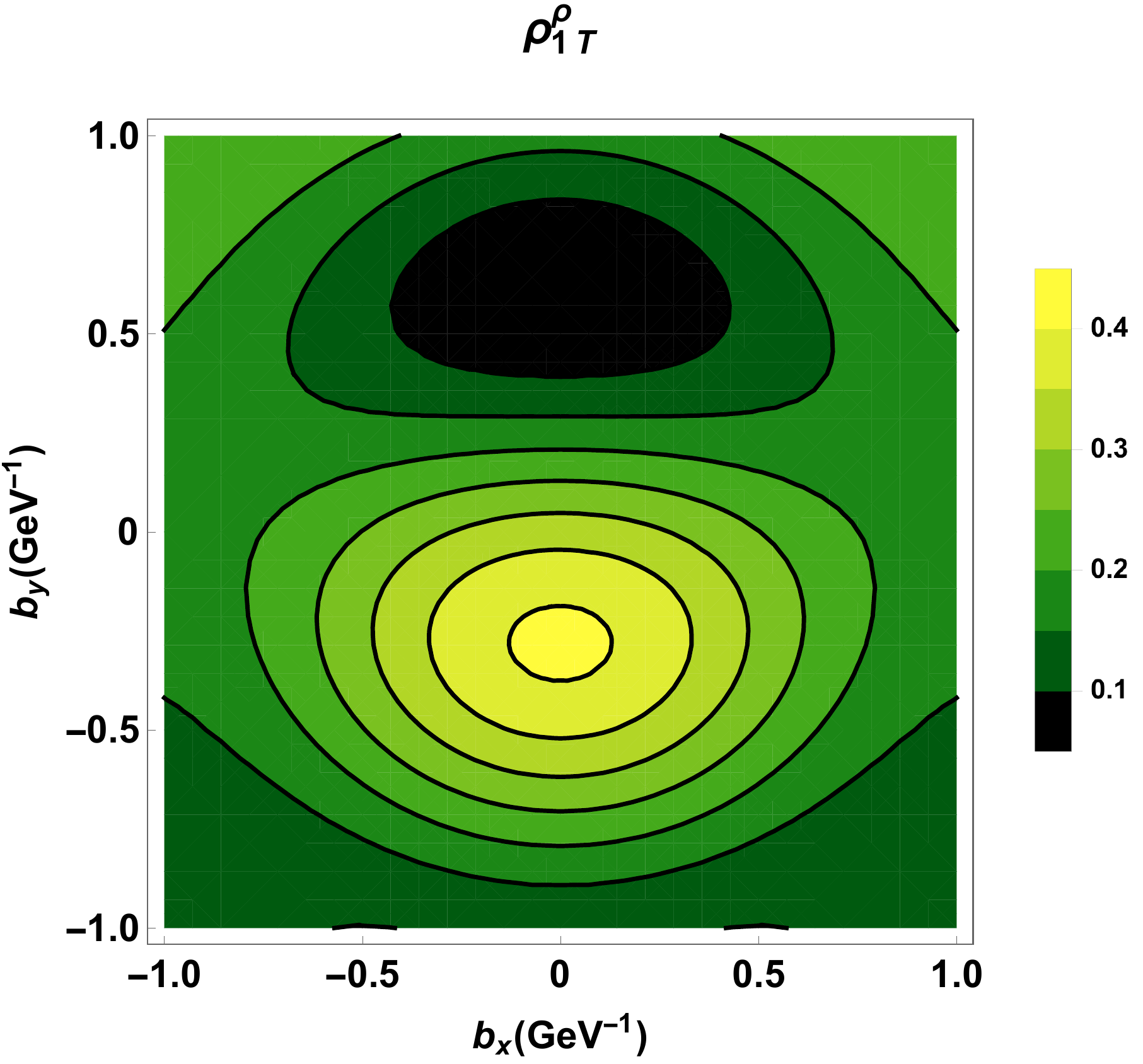}
    \hspace{0.1cm}
    \end{minipage}
\begin{minipage}[c]{0.98\textwidth}
    \small{(c)}\includegraphics[width=7cm]{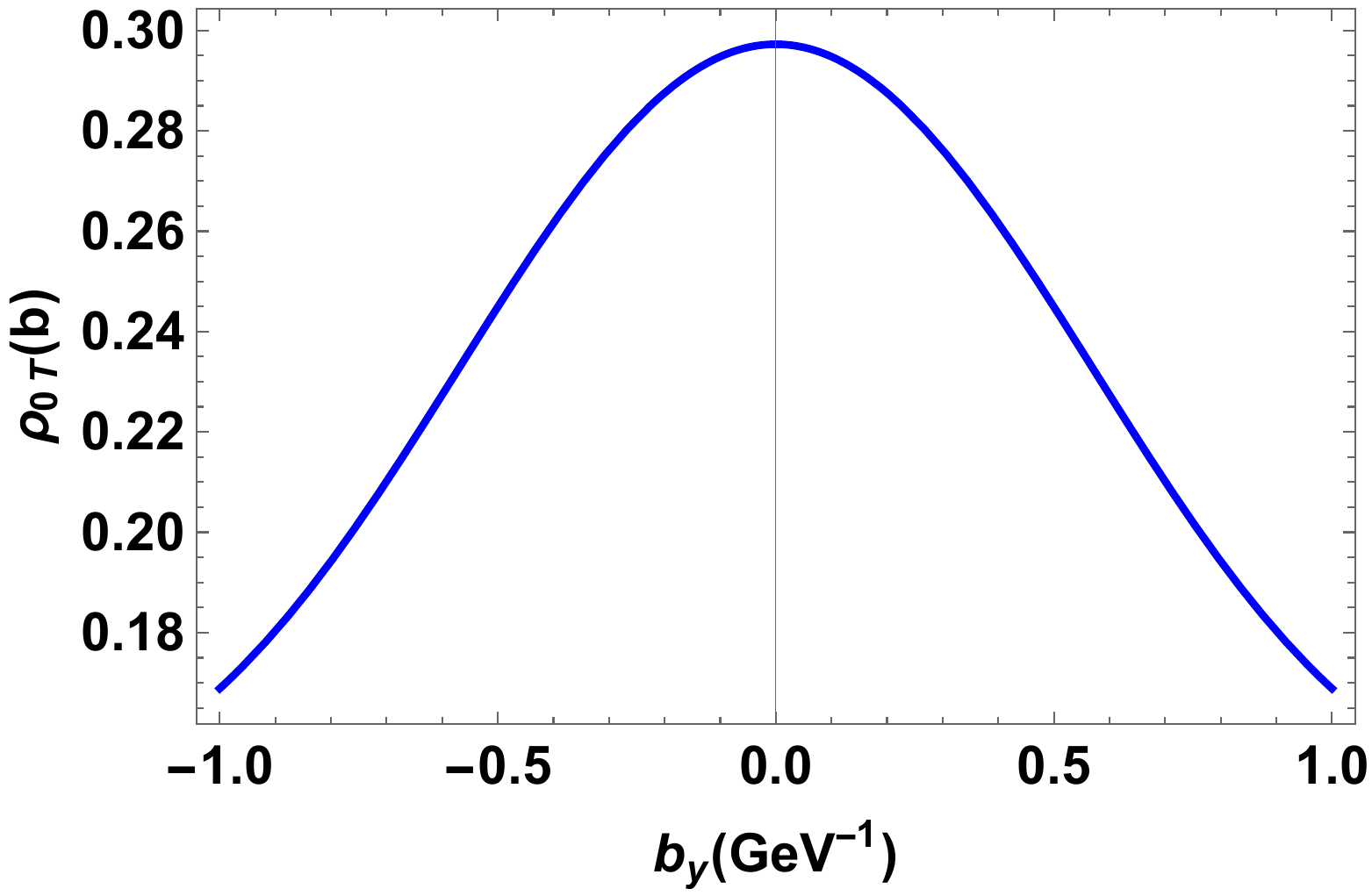}
    \hspace{0.1cm}
    \small{(d)}\includegraphics[width=7cm]{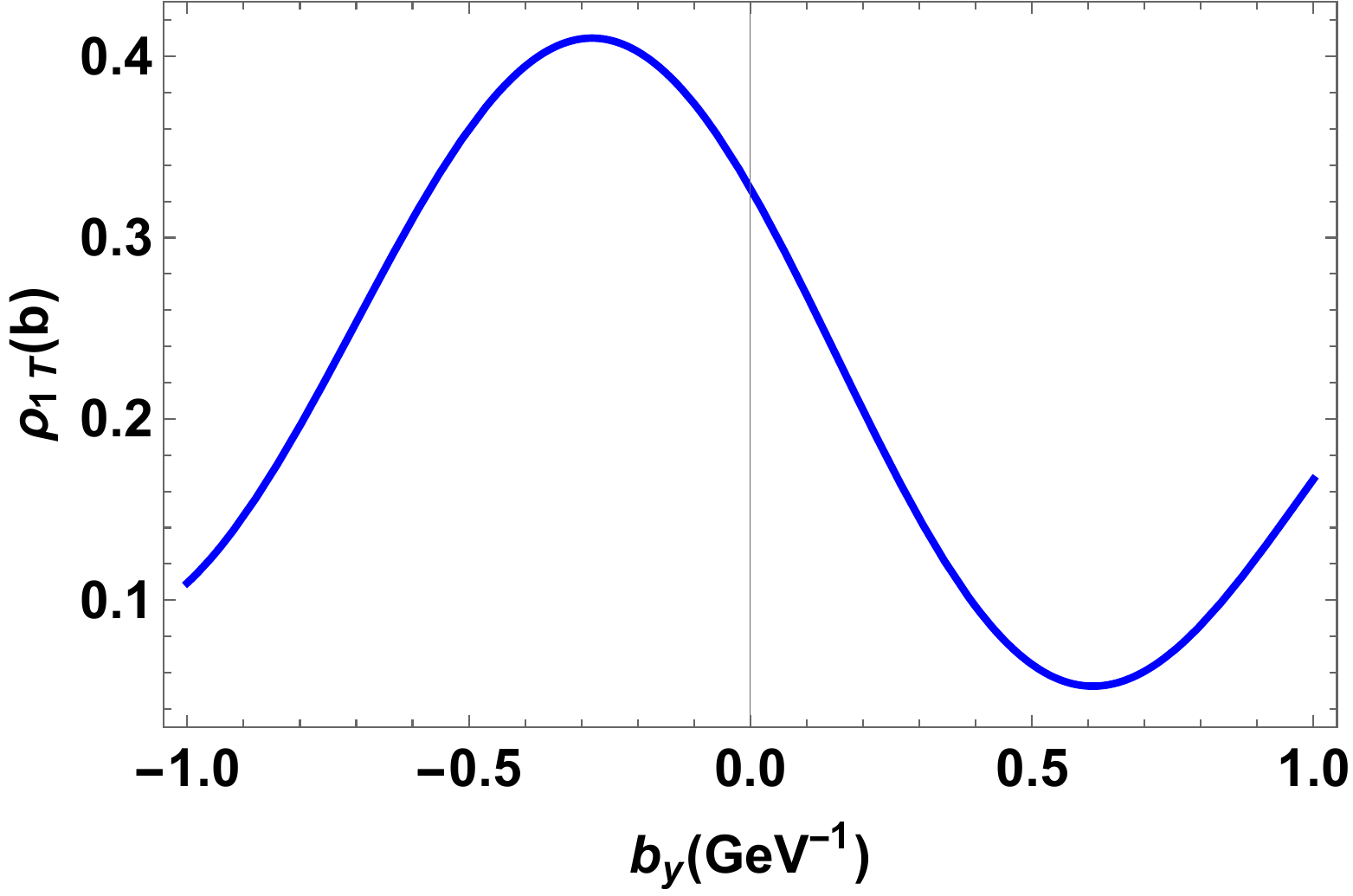}
    \hspace{0.1cm}
    \end{minipage}
\caption{(Color online) Plot for (a) monopole and (b) dipole contributions towards $\rho_0^T$ and correspondingly two-dimensional plots in (c) and (d) for $\rho$ meson in LFQM.}
\label{transverse-polarized-rho}
\end{figure*}
\section{Transverse charge densities for $\rho$ meson}
\label{sec:3}
\begin{figure*}
\begin{minipage}[c]{0.98\textwidth}
    \small{(a)}\includegraphics[width=7cm]{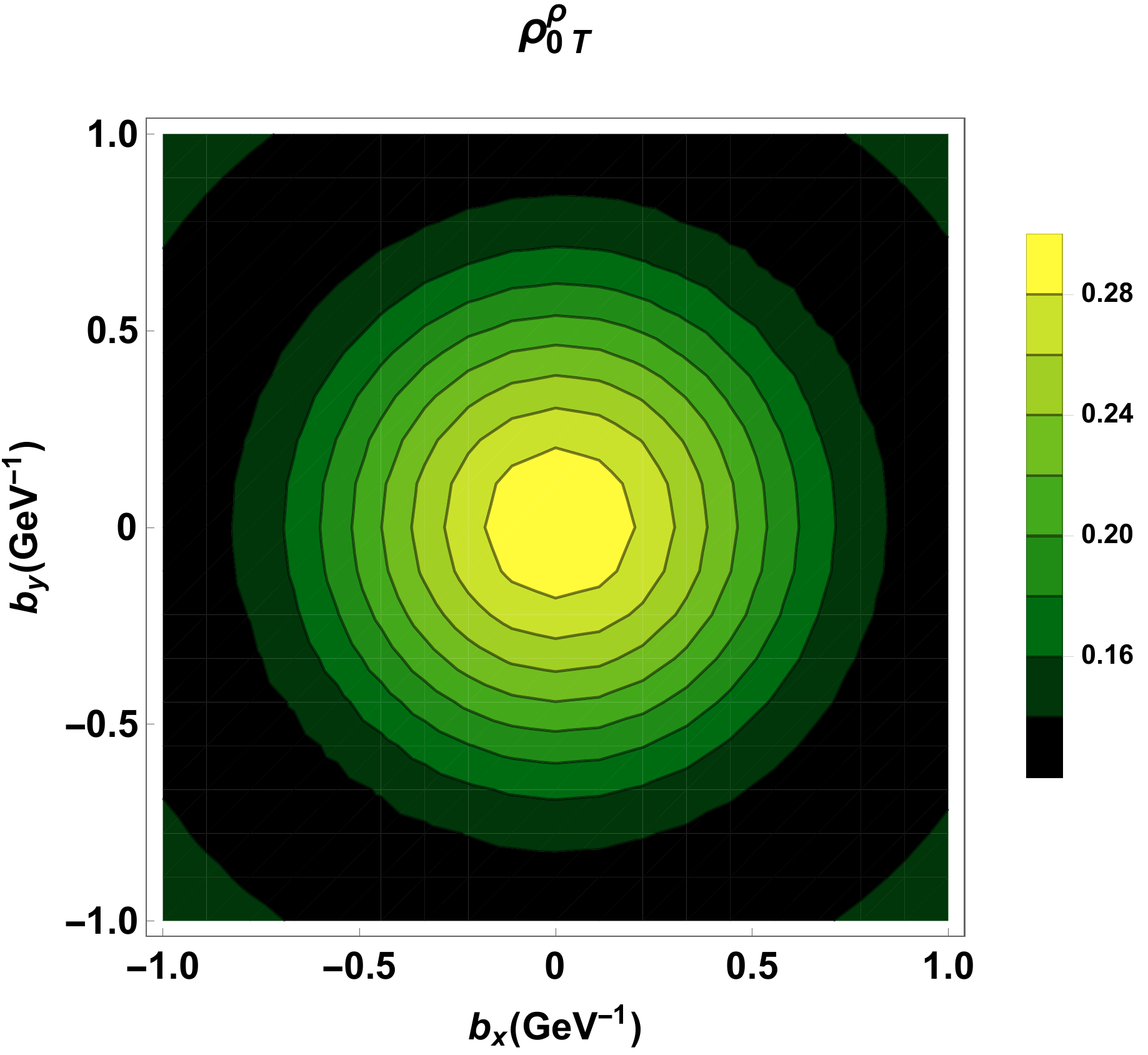}
    \hspace{0.1cm} 
    \small{(b)}\includegraphics[width=7cm]{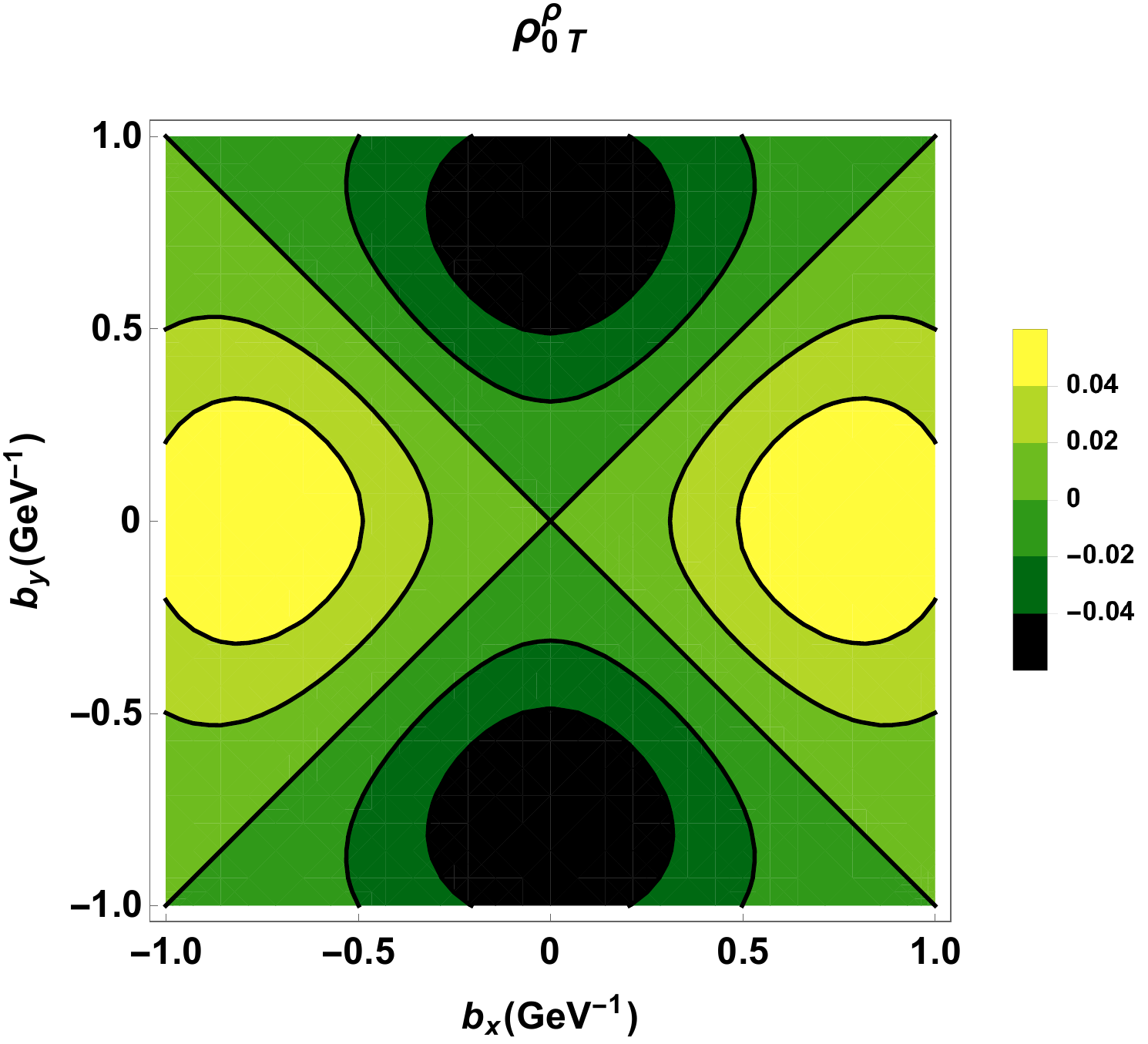}
    \hspace{0.1cm} 
    \end{minipage}
\caption{(Color online) Plot for (a) monopole and (b) quadrupole contributions towards $\rho_0^T$ for $\rho$ meson in LFQM.}
\label{rho0-mono-quad}
\end{figure*}
The charge density in transverse plane as a standard interpretation can be obtained by two-dimensional Fourier transform of form factor. In a true relativistic picture, form factors are Lorentz invariant and depend upon the Lorentz invariant quantities only. In the present work, we have extracted the transverse densities from the helicity matrix elements $G^+_{\Lambda' \Lambda}$ which is further obtained from the certain combinations of charge $G_C$, magnetic $G_M$ and quadrupole $G_Q$ form factors calculated in 
Breit frame within LFQM. It is to be noted that helicity matrix element is frame dependent which in turns reflect that transverse charge densities obtained from them are frame-dependent  and they are not Lorentz invariant. 
 This circumstance is also cleared from Ref. \cite{Mondal:2017lph}. The transverse charge densities for $\rho$ meson with light-front helicity state $\Lambda= \pm 1, 0$ are given as \cite{Carlson:2007xd,Miller:2007uy}
\begin{eqnarray}
\rho_\Lambda^\rho(b)&=& \int \frac{d^2 {\bf q}_\perp}{(2 \pi)^2} e^{-i {\bf q}_\perp \cdot {\bf b}_\perp} G_{\Lambda \Lambda}^{+}(Q^2),\nonumber\\
&=& \int_{0}^{\infty} \frac{dQ \ Q}{2 \pi} J_0(Q b) G_{\Lambda \Lambda}^{+}(Q^2),
\end{eqnarray}
where $G_{\Lambda' \Lambda}^{+}(Q^2)$, is the matrix element obtained from the electromagnetic current $J^+(0)$ sandwich between two $\rho$ meson states \cite{Carlson:2008zc}
\begin{eqnarray}
\langle P^+, \frac{{\bf q}_\perp}{2}, \Lambda' | J^+| P^+, -\frac{{\bf q}_\perp}{2}, \Lambda \rangle = 2 P^+ e^{i(\Lambda - \Lambda') \phi_q} G^+_{\Lambda' \Lambda}(Q^2),
\end{eqnarray}
here $\Lambda= \pm 1,0$ ($\Lambda'= \pm 1,0$) are the light-front helicity for initial (final) $\rho$ meson state. We also define ${\bf q}_\perp = Q (\cos \phi \ \hat{x} + \sin \phi \ \hat{y})$ and impact-parameter ${\bf b}_\perp= b (\cos \phi \ \hat{x} + \sin \phi \ \hat{y})$ in the transverse plane gives the position of parton from center of position space for the $\rho$ meson. Further, one can define the helicity- conserving matrix elements ($G_{11}^{+}$, $G_{00}^{+}$) and helicity non-conserving matrix elements ($G_{0+}^{+}$, $G_{-+}^{+}$) respectively, in terms of $G_C, G_M$ and $G_Q$ as \cite{Carlson:2008zc}:
\begin{eqnarray}
G_{++}^{+}&=& \frac{1}{1+\kappa}\bigg[G_C + G_M+ \frac{\kappa}{3} G_Q\bigg],\nonumber\\
G_{00}^{+}&=& \frac{1}{1+\kappa}\bigg[(1-\kappa)G_C + 2 \kappa G_M-  \frac{2 \kappa}{3}(1+2 \kappa) G_Q\bigg], \nonumber\\
G_{0+}^{+}&=& - \frac{\sqrt{2 \kappa}}{1+\kappa} \bigg[G_C - \frac{1}{2} (1-\kappa) G_M+ \frac{\kappa}{3} G_Q \bigg],\nonumber\\
G_{-+}^{+}&=& \frac{\kappa}{1+\kappa} \bigg[G_C - G_M - \bigg(1+\frac{2 \kappa}{3}\bigg) G_Q \bigg],
\end{eqnarray}
where again $\kappa= \frac{Q^2}{4 M_v^2}$. One should keep in mind that $\rho$ meson helicity matrix element $G_{0+}^{+}$ will give the dipole pattern whereas $G_{-+}^{+}$  will give the quadrupole pattern after getting transform into impact-parameter space in terms of transverse density. The reason of dipole and quadrupole pattern is due to angular dependencies which are associated with the one unit of helicity-flip ($G_{0+}^{+}$) matrix element and two unit of helicity-flip ($G_{-+}^{+}$) matrix element which can be seen in Eq. \ref{spin-1-trans} respectively. In Fig. \ref{helicity-ffs}(a) and (b), we present the results for helicity-conserving matrix elements $G_{00}^{+}$ and $G_{++}^{+}$ respectively whereas results for helicity non-conserving matrix elements ($G_{0+}^{+}$ and $G_{-+}^{+}$) are shown in Fig. \ref{helicity-ffs}(c) and (d). We present the results for the charge density for unpolarized $\rho$ meson in Fig. \ref{unpolarized-rho-meson} (a) and (b). We observe that the distribution is axially symmetric or one can say monopole in nature for both the cases and peak at the center of impact-parameter space. In Fig. \ref{unpolarized-rho-meson} (c) and (d), we present the results in 2D plots. We also consider the transversely polarized $\rho$ meson state which provides information about dipole and quadrupole moments. Transverse charge density for transversely polarized $\rho$ meson can be defined as 
\begin{eqnarray}
\rho^\rho_{s_\perp T}({\bf b}_\perp)&=& \int \frac{d^2 {\bf q}_\perp}{(2 \pi)^2} e^{-i {\bf q}_\perp \cdot {\bf b}_\perp} \frac{1}{2 P^+} \langle P^+, \frac{{\bf q}_\perp}{2}, s_\perp | J^+ | P^+, -\frac{{\bf q}_\perp}{2}, s_\perp \rangle,
\end{eqnarray}
where $s_\perp$ is the $\rho$ meson transverse spin projection along the transverse polarization direction $S_\perp= \cos \phi \ \hat{x}+ \sin \phi \ \hat{y}$ and for $s_\perp = 0, 1 $, tranverse charge densities can be written as \cite{Carlson:2008zc}
\bea
\rho_{0T}(b) &=& \int_{0}^{\infty} \frac{ dQ \ Q}{2\pi} \bigg[J_0(b Q) G_{++}^{+} + \cos2(\phi_b - \phi_s) J_2(b Q) G_{+-}^{+}\bigg],
\label{spin-0-trans}
\eea
\bea
\rho_{1T}(b)&=& \int_{0}^{\infty} \frac{ dQ \ Q}{2\pi} \bigg[\frac{J_0(b Q)}{2} (G_{++}^{+} + G_{00}^{+} ) + \sin(\phi_b - \phi_s) J_1(b Q) \sqrt{2} G_{0+}^{+} - \nonumber\\
&&\cos 2(\phi_b - \phi_s) J_2(b Q) \frac{G_{-+}^{+}}{2}  \bigg].
\label{spin-1-trans}
\eea
\begin{figure*}
\begin{minipage}[c]{0.98\textwidth}
    \small{(a)}\includegraphics[width=7cm]{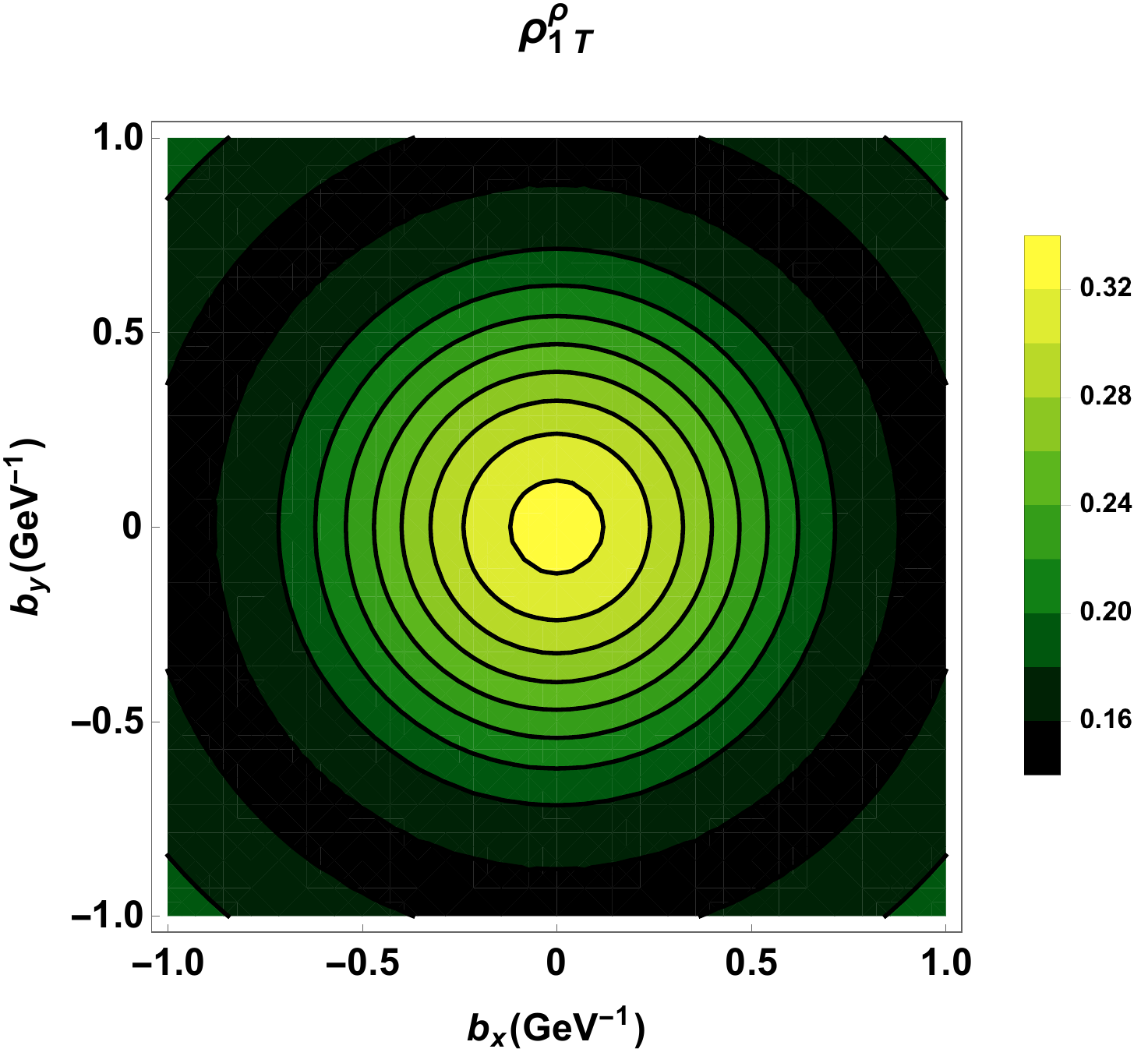}
    \hspace{0.1cm}
        \small{(b)}\includegraphics[width=7cm]{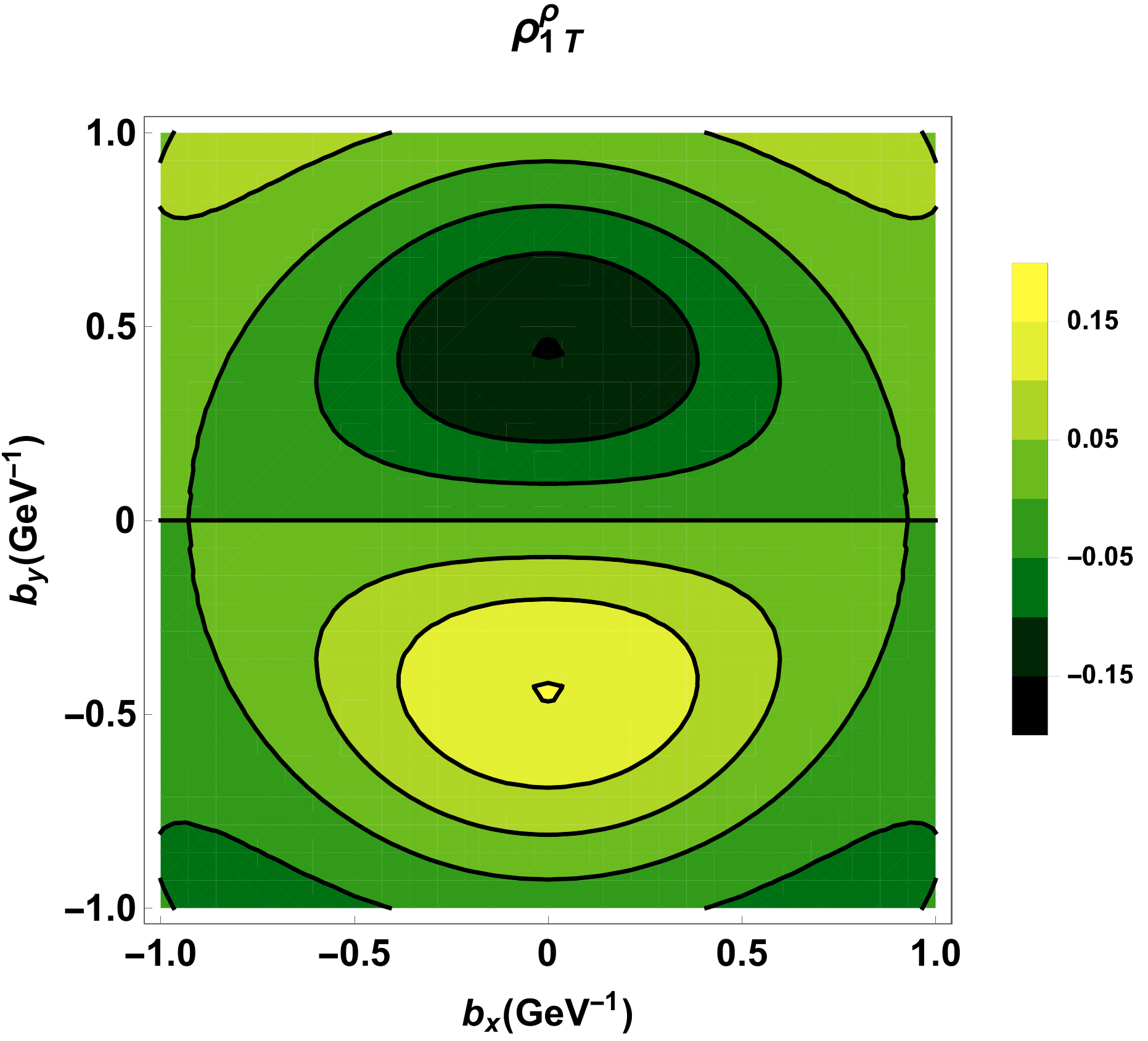}
        \hspace{0.1cm}
    \small{(c)}\includegraphics[width=7cm]{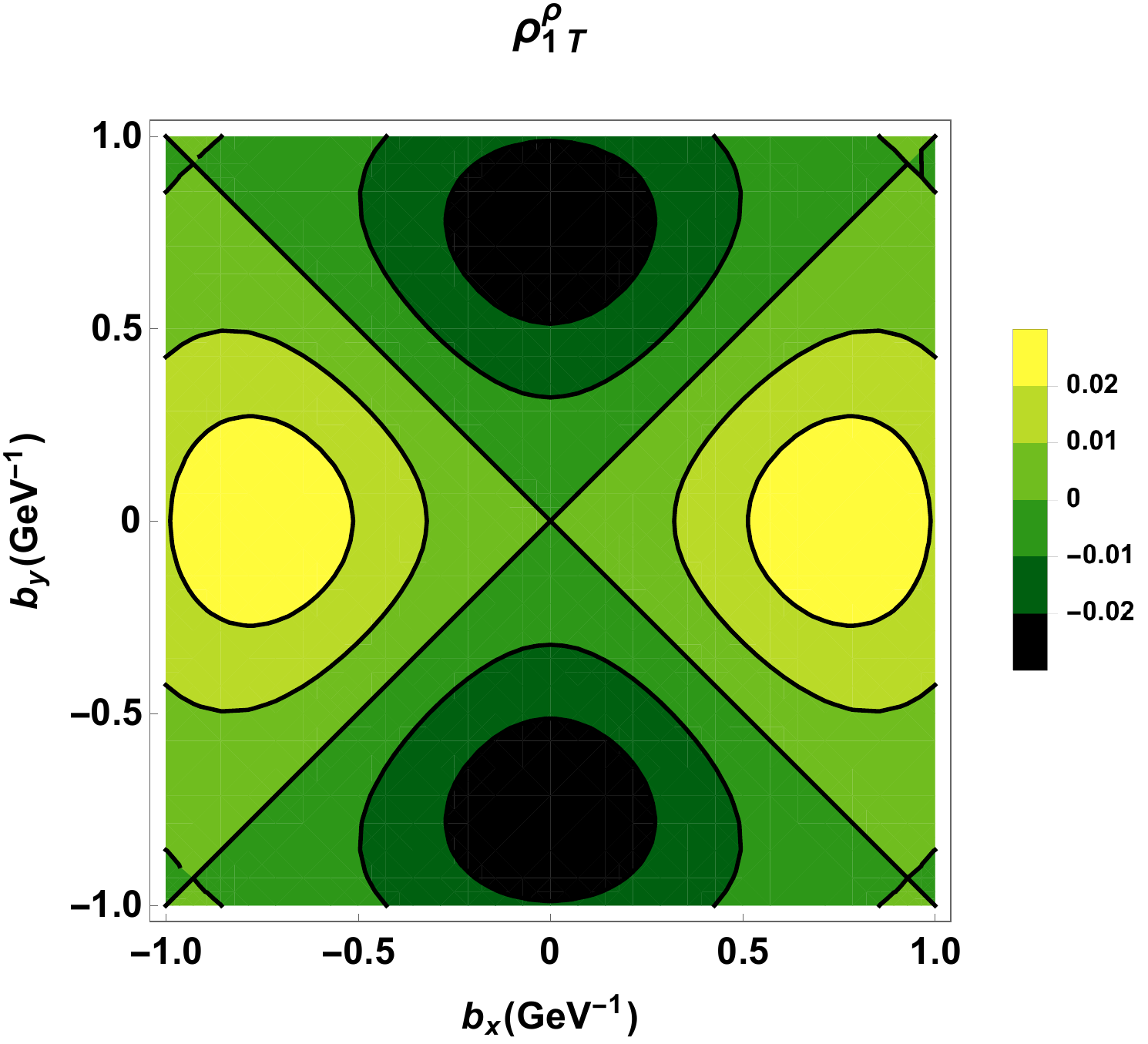}
        \hspace{0.1cm}
    \end{minipage}
\caption{(Color online) Plots for (a) monopole, (b) dipole and (c) quadrupole contributions towards $\rho_1^T$ for $\rho$ meson in LFQM.}
\label{polarized-dipole-quadrupole}
\end{figure*}
Without loss of generality, we choose $\phi_s=0$. In $\rho_{0T}$, we are getting contributions from two helicity matrix elements $G_{++}^{+}$ and $G_{-+}^{+}$. The former contains no helicity flip but later includes a net shift of two units of helicity i.e., $-1 \rightarrow +1$.  However in $\rho_{1T}$, contributions are received from $G_{0+}^{+}$ (total helicity flip $ 0 \rightarrow 1$) and again $G_{-+}^{+}$. In Fig. \ref{transverse-polarized-rho} (a) and (c), we present the results for  $\rho_{0T}$ which show that distribution gets stretched along $\hat{y}$ axis after getting contribution from $G_{+-}^{+}$ but their is no overall shift of the peak in the charge density. However, in Fig. \ref{transverse-polarized-rho} (b) and (d) we consider the $\rho$ meson to be polarized along $\hat{x}$-direction. We found that $\rho_{1T}$ received contributions from $G_{0+}^{+}$ and $G_{-+}^{+}$. Due to significant contributions received from the dipole and quadrupole term the distribution gets distorted from the center of distribution. In Fig.  \ref{rho0-mono-quad} (a) and (b), we show the results for the monopole and quadrupole contributions of $\rho_{0T}$. We also present the results for monopole, dipole and quadrupole contributions for $\rho_{1T}$ in Fig. \ref{polarized-dipole-quadrupole}. It is to be noted that dipole and quadrupole patterns in charge densities are due to anomalous magnetic moment coming from second term of Eq. \ref{spin-1-trans} which produced an electric dipole moment in $\hat{y}$ direction and anomalous quadrupole moment comes from the third term of Eq. \ref{spin-1-trans}. It is also noted that signs of quadrupole contributions are same in $\rho_{0T}$ and $\rho_{1T}$, which stretch the distribution for $\rho_T(0)$ and distort the distribution significantly for $\rho_{1T}$ in $\hat{y}$ direction. 
\begin{figure*}
\begin{minipage}[c]{0.98\textwidth}
    \small{(a)}\includegraphics[width=7cm]{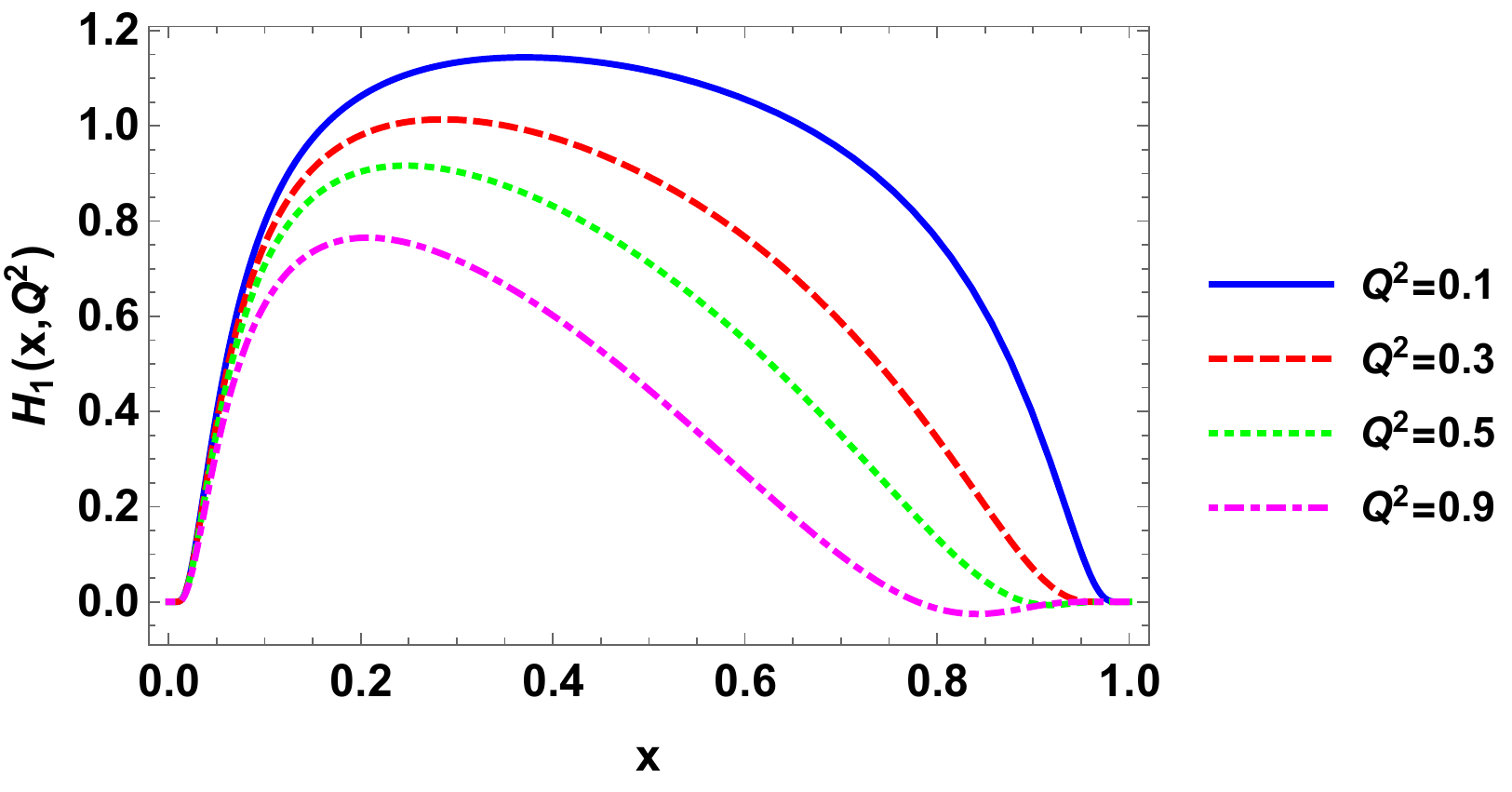}\hspace{.1cm}
    \small{(b)}\includegraphics[width=7cm]{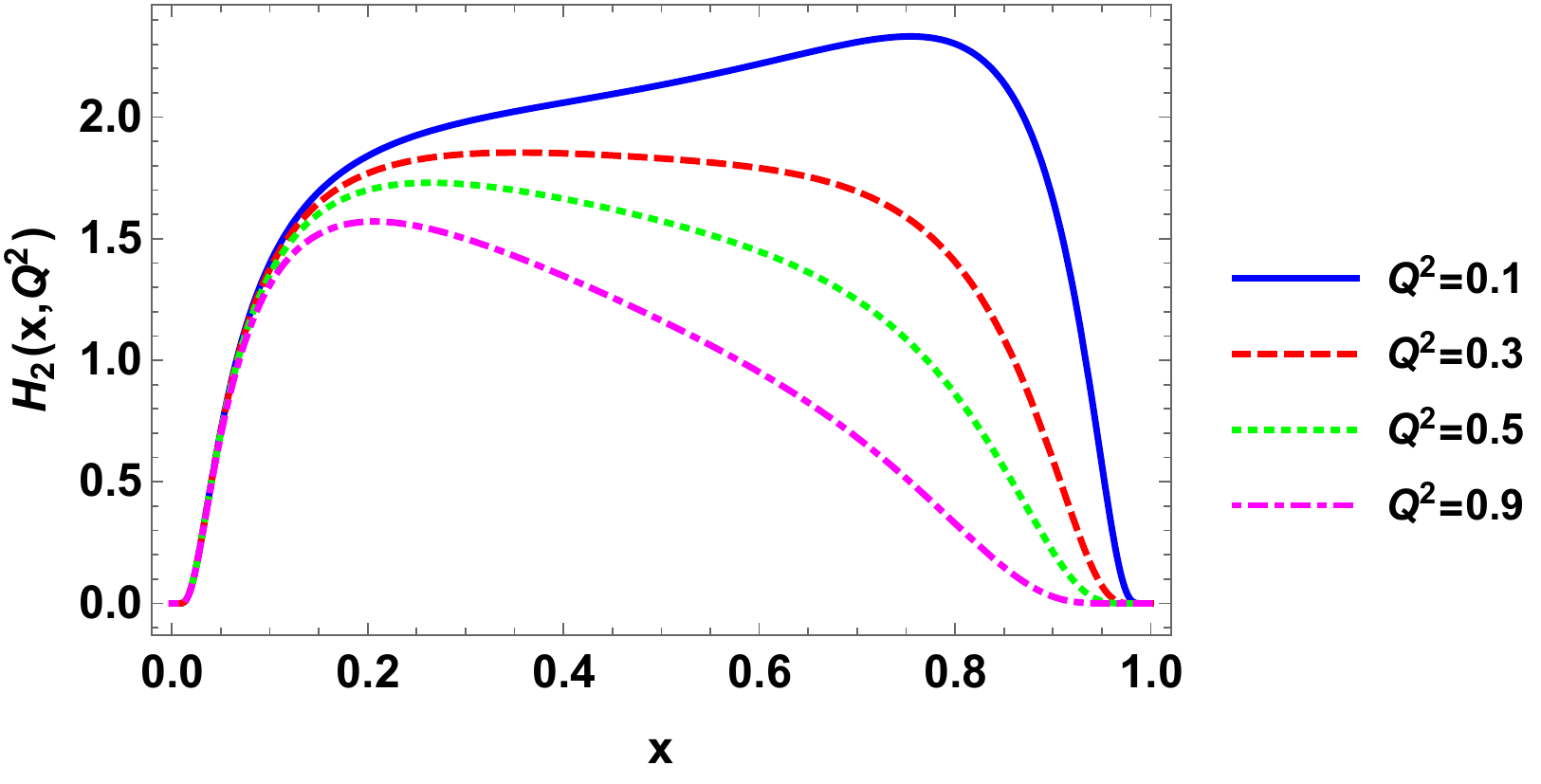}\hspace{.1cm}
        \small{(c)}\includegraphics[width=7cm]{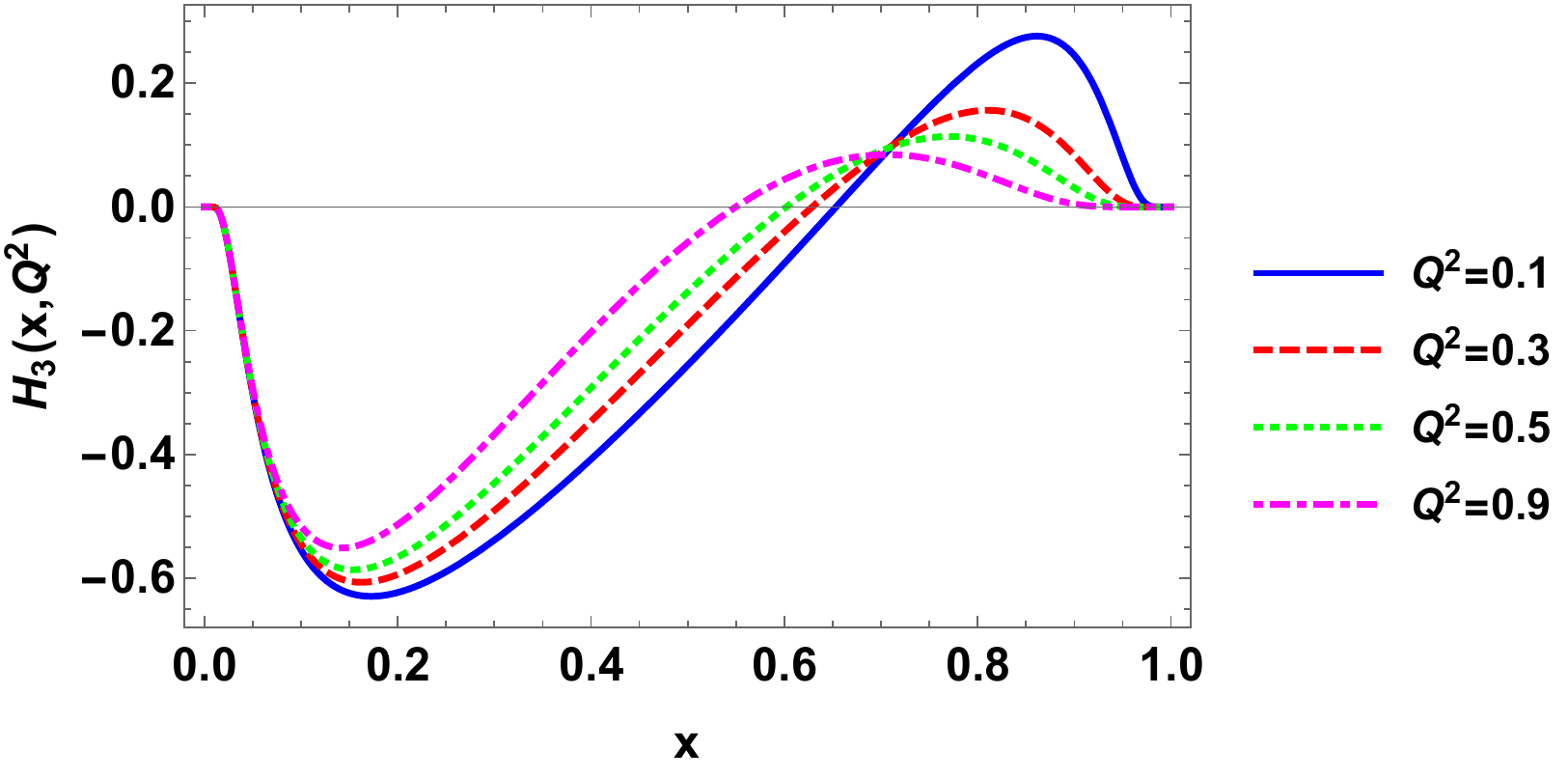}\hspace{.1cm}
            \end{minipage}
\caption{(Color online) GPDs of the $\rho$ meson evaluated in LFQM(a) $H_1(x,Q^2)$ (b)$H_2(x,Q^2)$ and (c) $H_3(x,Q^2)$ with fixed values of $Q^2$.}
\label{gpds-plot}
\end{figure*}
\section{Generalized parton distributions for the $\rho$ meson}
\label{sec:4}
Generalized parton distributions encode the three-dimensional structure of the hadrons. One can predict the physical form factors from the first moment of GPDs i.e.,
\begin{equation}
G_z(Q^2)= \int dx H_i(x,Q^2),
\label{gpds}
\end{equation}
where $z=C,M$ and $Q$ and $i=1,2$ and $3$ respectively. The integrals over $H_4$ and $H_5$ vanishes due to constraints of Lorentz invariance and time reversal. GPDs for the $\rho$ meson can be defined by the correlator function as \cite{Berger:2001zb}
\begin{eqnarray}
V_{\lambda' \lambda}&=& \frac{1}{2} \int \frac{d\omega}{2\pi} e^{i x (P z)} \langle P', \lambda'| \bar{q}(-z/2) \slashed{n} q(z/2)| P, \lambda \rangle|_{z= \omega n}, \nonumber\\
&=& \sum_i \epsilon^{' * \nu} V_{\nu \mu}^{(i)} \epsilon^\mu H_i(x, \xi, t),
\end{eqnarray}
where $\epsilon= \epsilon(p, \lambda)$ or $[\epsilon'= \epsilon(p',\lambda')]$ and $\lambda=0$ ($\lambda'=\pm 1$) are the initial (final) polarization vector and helicity respectively. GPDs of the $\rho$ meson are defined as
\begin{eqnarray}
V_{\lambda' \lambda}&=& - (\epsilon^{*'} \cdot \epsilon) H_1 + \frac{(\epsilon \cdot n)(\epsilon^{*'} \cdot P)+(\epsilon^{*'} \cdot n) (\epsilon \cdot P)}{P \cdot n} H_2- \frac{2 (\epsilon \cdot n)(\epsilon^{*'} \cdot n)}{M_v^2} H_3 \nonumber\\
&+&\frac{(\epsilon \cdot n)(\epsilon^{*'} \cdot P)-(\epsilon^{*'} \cdot n) (\epsilon \cdot P)}{P \cdot n} H_4+ \bigg[\frac{M_v^2 (\epsilon \cdot n)(\epsilon^{*'} \cdot n)}{(P \cdot n)^2} + \frac{1}{3} (\epsilon^{*'} \epsilon)\bigg] H_5,
\end{eqnarray}
\begin{figure*}
\begin{minipage}[c]{0.98\textwidth}
    \small{(a)}\includegraphics[width=7cm]{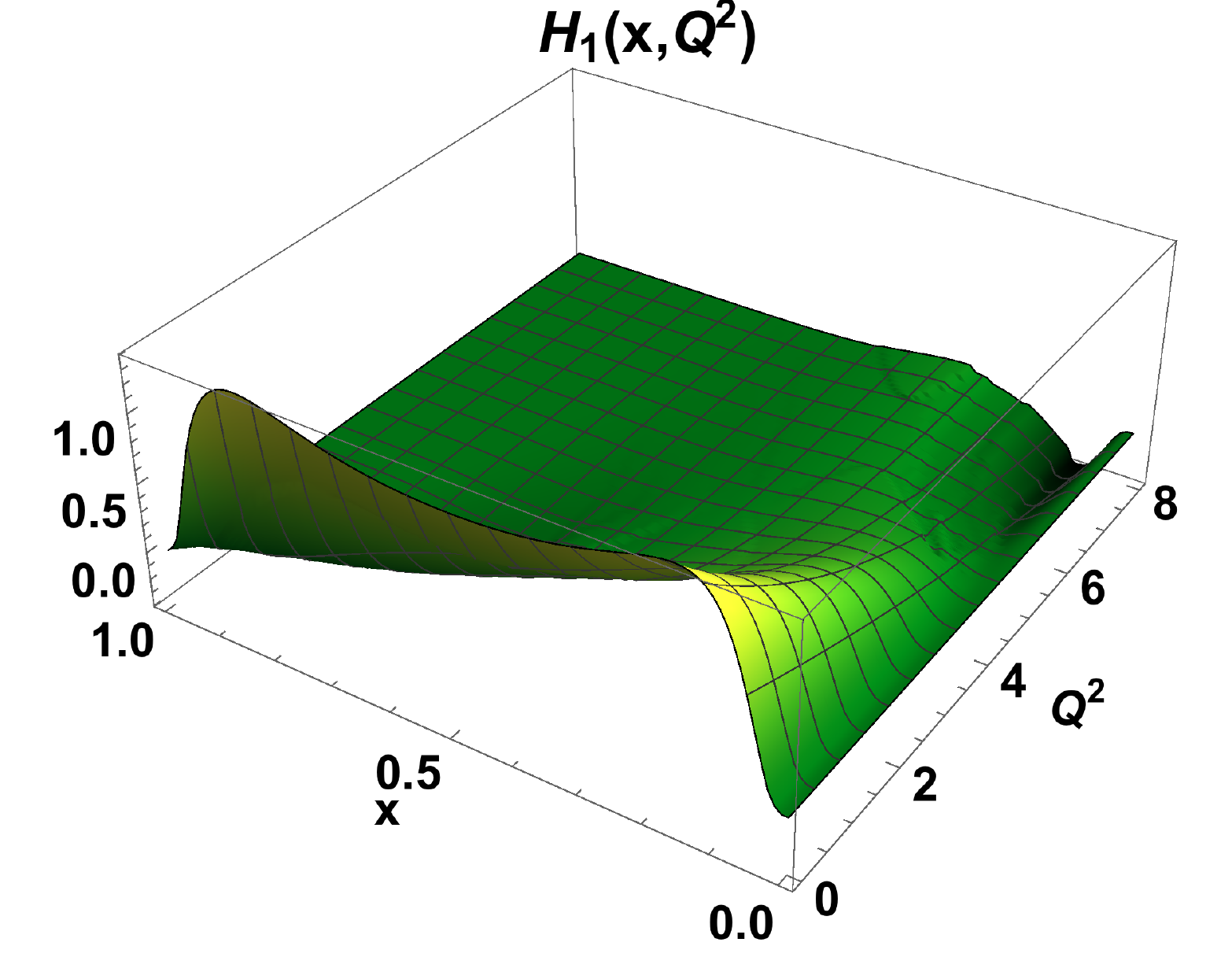}\hspace{.1cm}
    \small{(b)}\includegraphics[width=7cm]{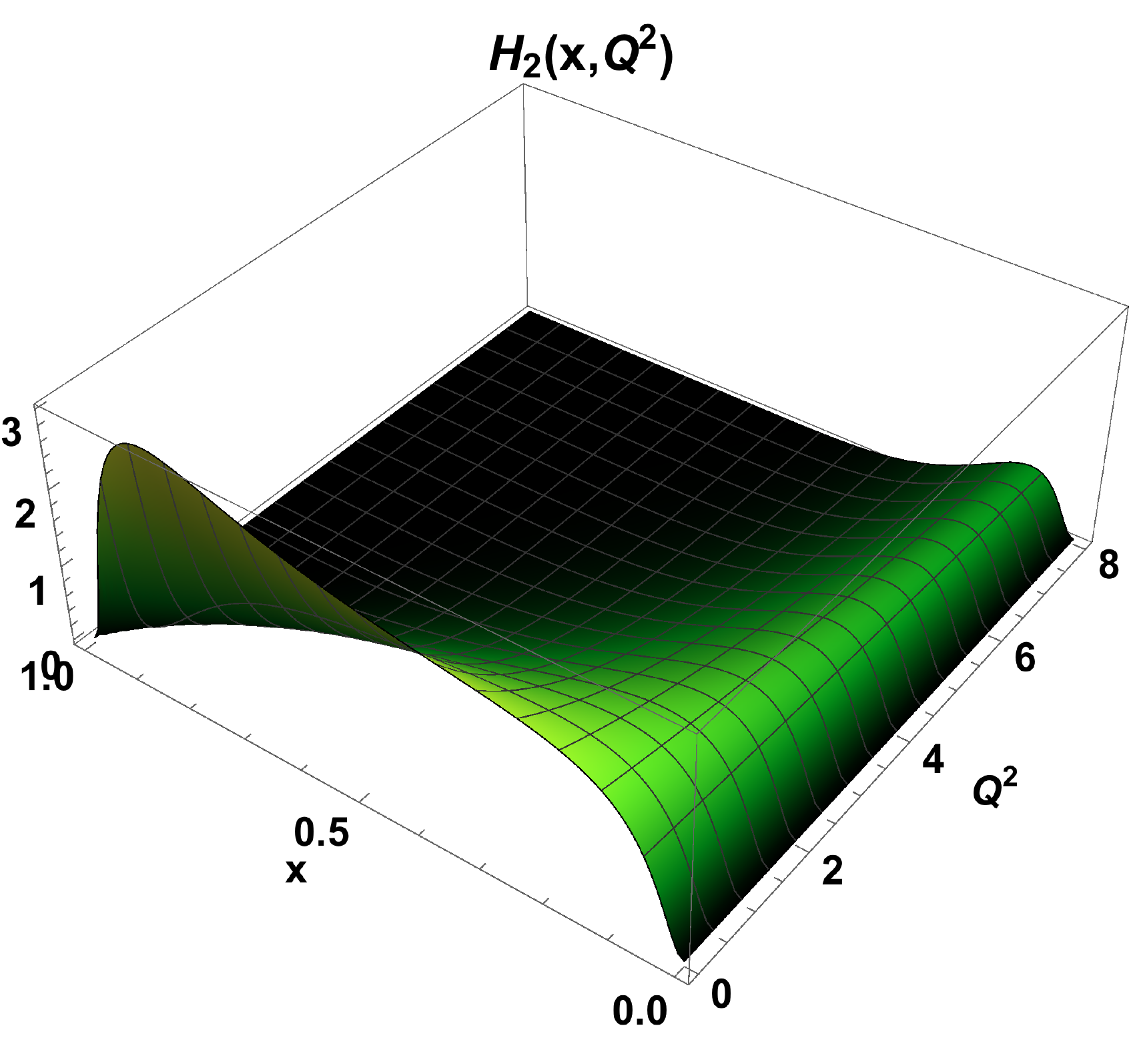}\hspace{.1cm}
    \small{(c)}\includegraphics[width=7cm]{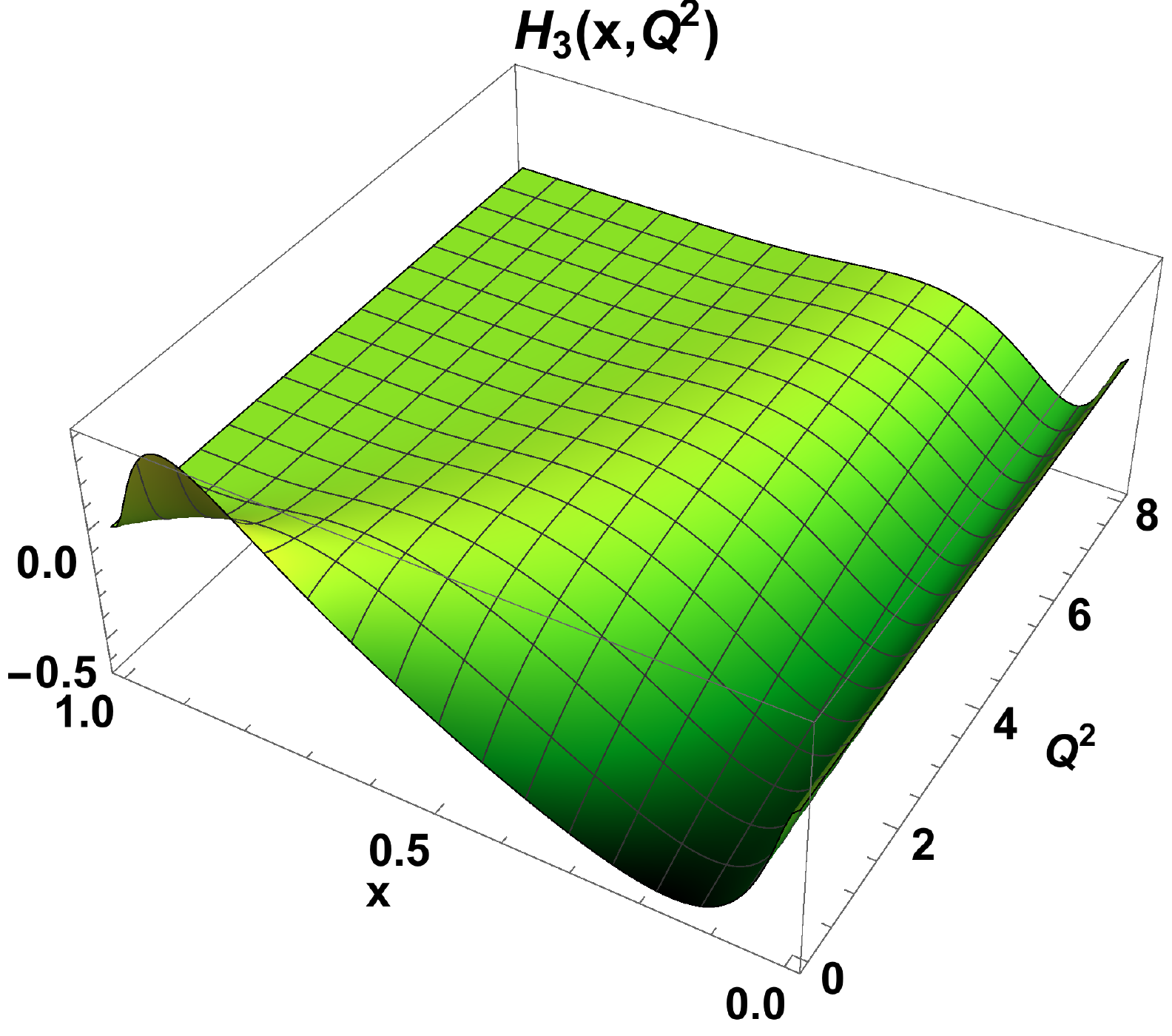}\hspace{.1cm}
\end{minipage}
\caption{(Color online) 3D representation of GPDs for the $\rho$ meson evaluated in LFQM (a) $H_1(x,Q^2)$ (b)$H_2(x,Q^2)$ and (c) $H_3(x,Q^2)$.}
\label{gpds-3d}
\end{figure*}
where GPDs $H_i (i= 1 \ \text{to} \ 5)$ are functions of $x$, $\xi$ and $t$. These three variables represent the  fraction of plus components, skewness and square of the momentum transfer during the process, respectively. In the forward limit, GPDs reduce to ordinary parton distribution functions (PDFs). In order to obtain the GPDs of the $\rho$ meson, we have compared Eq. \ref{physical-ffs} and \ref{gpds}. Corresponding to each physical FFs i.e. $G_C, G_M$ and $G_Q$, we can define the GPDs $H_1, H_2$ and $H_3$ respectively. In Fig. \ref{gpds-plot}, we have presented the results for $\rho$ meson GPDs for fixed values of $Q^2$ with respect to $x$. We found that for low value of $Q^2$, the magnitude of peak is maximum at large value of $x$ but as we increase the $Q^2$, magnitude of peak decreases and also shifted towards lower value of $x$. This shows that active quark in $\rho$ meson is dominating at lower value of $x$ i.e. most of the momentum carried by the active quark is at lower values of $x$. This fact is also evident from plots shown in Fig. \ref{gpds-3d}(a), (b) and (c) respectively. However in Fig. \ref{gpds-plot}(c), we have observed that GPDs appears to be negative which is due to presence of -ve sign outside of $G_Q$ form factor.
 
\begin{figure*}
\begin{minipage}[c]{0.98\textwidth}
    \small{(a)}\includegraphics[width=7cm]{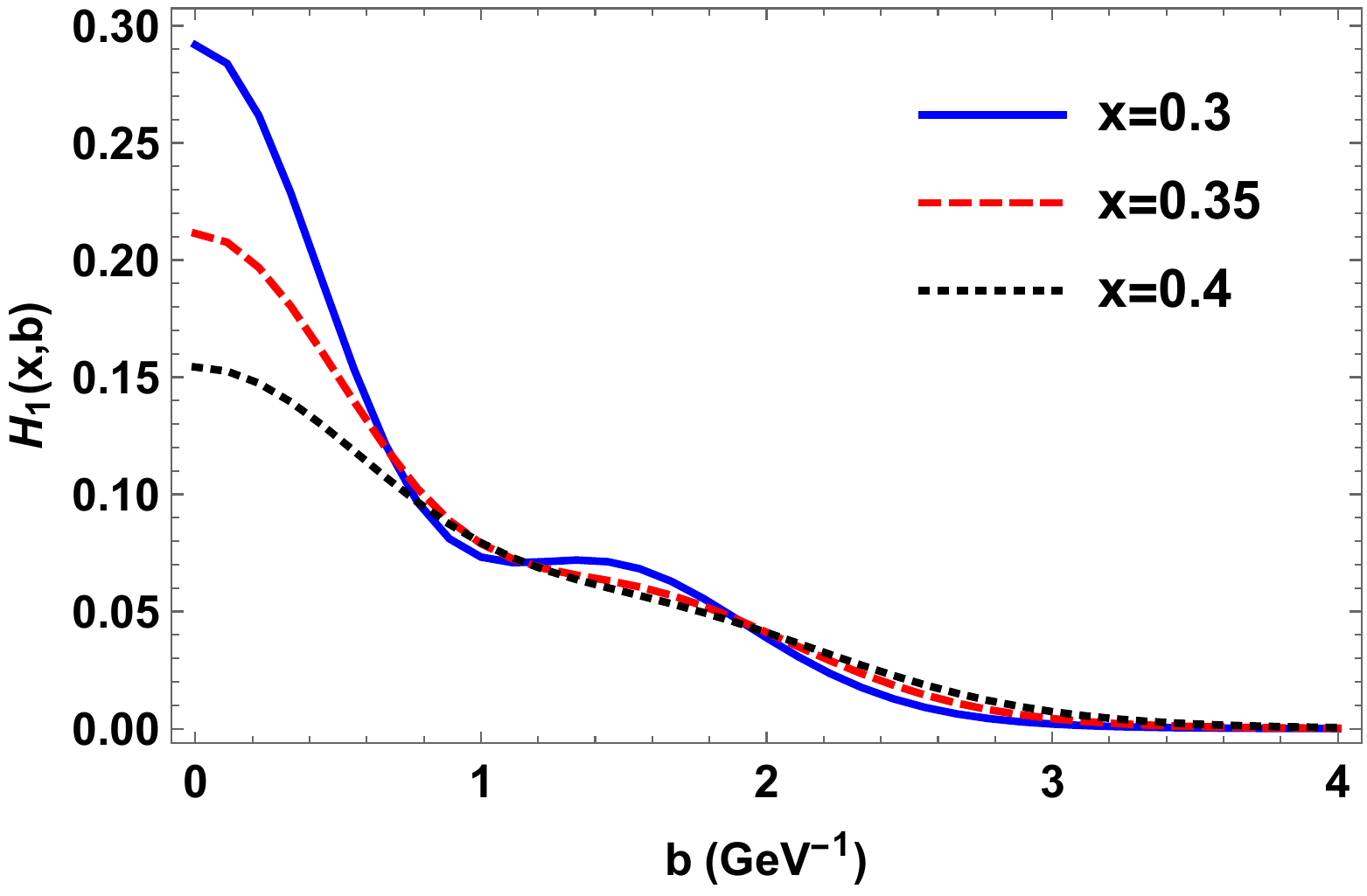}\hspace{0.1cm}
  \small{(b)}\includegraphics[width=7cm]{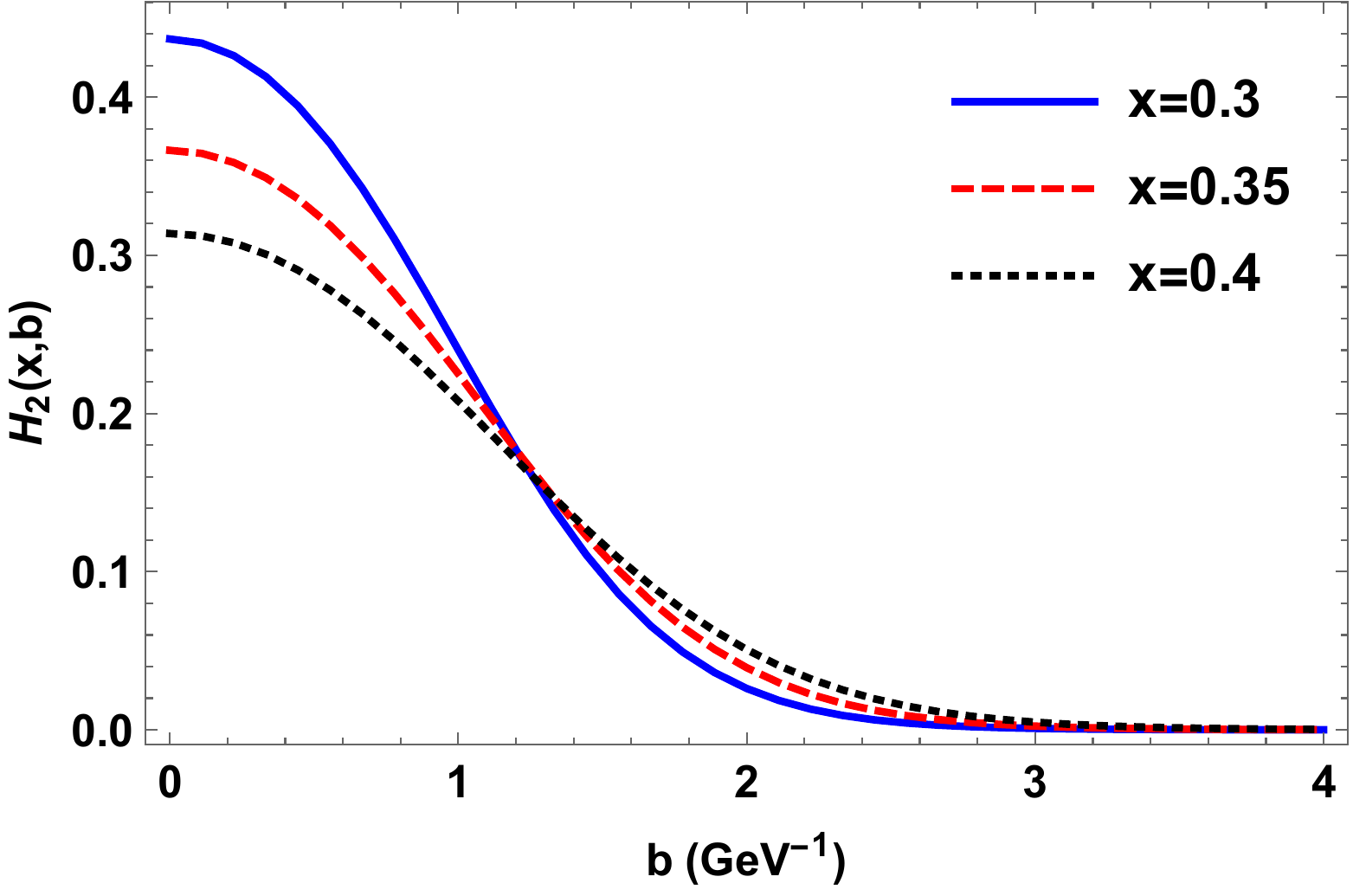}\hspace{0.1cm}
    \small{(c)}\includegraphics[width=7cm]{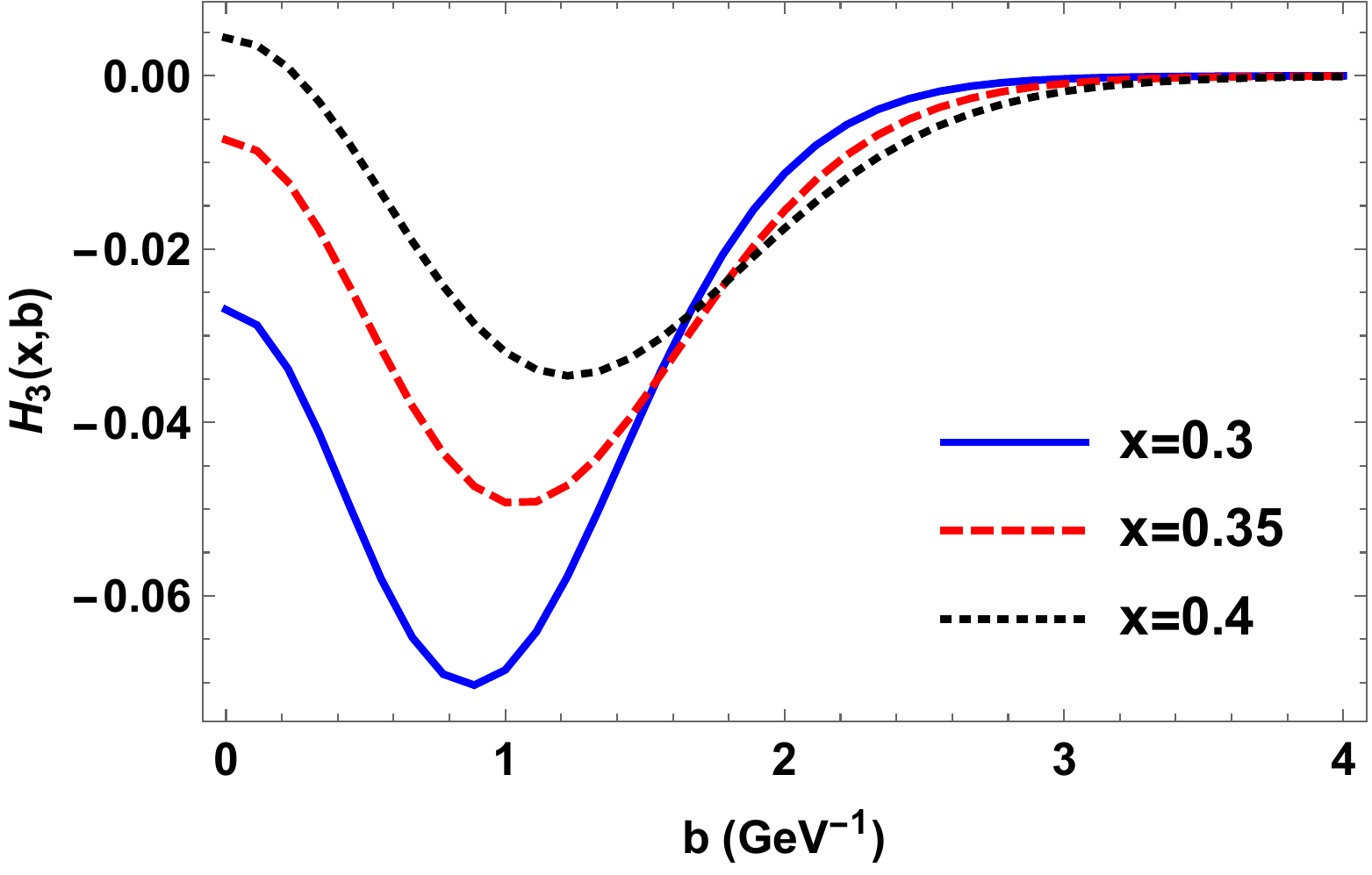}\hspace{0.1cm}
            \end{minipage}
\caption{(Color online)IPDPDF of the $\rho$ meson evaluated in light-front quark model (a) $H_1(x,b)$ (b)$H_2(x,b)$ and (c) $H_3(x,b)$ with fixed values of $x$.}
\label{ipdpdf}
\end{figure*}
\section{Impact Parameter Dependent Parton Distribution Function}
\label{sec:5}
In this section, we have discussed the ipdpdfs for the $\rho$ meson. The FT with respect to the momentum transfer $Q$ gives the GPDs in impact-parameter space. We introduce $b_\perp$ conjugate to $Q$ giving
\bea
H(x, b) &= &\int \frac{d^2 q}{(2 \pi)^2} e^{- i {\bf q_\perp} \cdot {\bf b_\perp}} H(x,Q^2), \nonumber\\
&=& \int \frac{Q \ dQ}{ 2 \pi} J_{0}(Q b) H(x, Q^2).
\eea
We have shown the results for the ipdpdf for $\rho$ meson for fixed values of $x$ with respect to $b$ in Fig. \ref{ipdpdf}(a),(b) and (c) respectively. It is clear from Fig. \ref{ipdpdf}(a) that partons that have small impact parameter has maximum peak at $b=0$, however we also noticed that as we increase the values of $x$, the magnitude of the peak decreases this shows that more the longitudinal momentum fraction carried by the parton less they are located near the centre of the position space in $\rho$ meson. Similar observation has been made in Fig. \ref{ipdpdf}(b). However, in Fig. \ref{ipdpdf} (c) we have observed that for $H_3$ the distribution of partons is not exactly near the center of position space but somehow shifted near to $b= 1.0$ \ GeV$^{-1}$. 
\section{Conclusions}
\label{sec:6}
In the present work, we have discussed the transverse densities for unpolarized and transversely $\rho$ meson in LFQM. Using the results of the physical form factors we have calculated the helicity matrix elements. We found that charge density for longitudinally polarised $\rho$ meson is axially symmetric. Charge distributions for transversely polarised $\rho$ meson show a monopole pattern together with dipole and quadrupole patterns. The dipolar structure in the transversely polarised $\rho$ meson causes the distortion in the charge distribution for $\rho_{1T}$ whereas quadrupole structure stretch the distribution for $\rho_{0T}$. In addition to this, we have also calculated the GPDs for $\rho$ meson obtained from the electromagnetic FFs. We found that in $\rho$ meson, active quark appears to carried more momentum at lower values of $x$. Finally, we have investigated the ipdpdfs for $\rho$ meson.  Although results are not Lorentz invariant but they can be examine by the experimental data if available and still they can be tested by the experiments with the appropriate choice of lab frames. Present work motivates to calculate the GPDs for $\rho$ meson by considering the quark-antiquark-meson vertex in LFQM and such calculations are in progress. It will be also interesting to calculate the deuteron GPDs in LFQM.
\section{Acknowledgements}
N.K. thanks Chueng-Ryong Ji (NCSU) for insightful discussions. N.K. also thankful to Dipankar Chakrabarti (IIT Kanpur) for carefully reading the manuscript and providing valuable suggestions. N.K. acknowledge financial support received from Science and Engineering Research Board-a statutory board under Department of Science and Technology, Government of India (Grant No. PDF/2016/000722, Project -``Study of Three Dimensional Structure of the Nucleon in Light Front QCD") under National Post-Doctoral Fellowship.

\end{document}